Focused Review

# Recent advances in spin-orbit torques: Moving towards device applications

Rajagopalan Ramaswamy, Jong Min Lee, Kaiming Cai and Hyunsoo Yang

*Department of Electrical and Computer Engineering, National University of Singapore, 117576, Singapore*

The ability of spintronic devices to utilize an electric current for manipulating the magnetization has resulted in large-scale developments, such as, magnetic random access memories and boosted the spintronic research area. In this regard, over the last decade, magnetization manipulation using spin-orbit torque has been devoted a lot of research attention as it shows a great promise for future ultrafast and power efficient magnetic memories. In this review, we summarize the latest advancements in spin-orbit torque research and highlight some of the technical challenges for practical spin-orbit torque devices. We will first introduce the basic concepts and highlight the latest material choices for spin-orbit torque devices. Then, we will summarize the important advancements in the study of magnetization switching dynamics using spin-orbit torque, which are important from scientific as well as technological aspect. The final major section focuses on the concept of external assist field free spin-orbit torque switching which is a requirement for practical spin-orbit torque devices.

**TABLE OF CONTENTS**

## I. Introduction

In conventional electronics, only the charge degree of freedom of an election is utilized to construct devices. The electron also possesses a spin angular momentum which is closely associated with its magnetic moment. The field of spintronics, or spin-based electronics, exploits both charge and spin degrees of freedom of electrons to provide additional functionalities to conventional electronic devices such as non-volatility and reduced power consumption. While one of the first studies to observe the interaction between a charge current and magnetism dates back to 1857,[1] the growth of spintronics was largely boosted by the discoveries of giant magnetoresistance (GMR)[2,3] and tunnel magnetoresistance (TMR)[4-6], in which the electrical resistance of a material system can vary significantly depending on the orientation of magnetic moments in the ferromagnetic layers. GMR and TMR based magnetic read heads replaced the conventional inductive read heads and led to a large boost in the areal density of hard disk drives.[7] However, it should be noted that the magnetization manipulation in the GMR/TMR read heads was still carried out using an external magnetic field.

Another major application of the spintronics lies in the electrical manipulation of the magnetic moment of a ferromagnet (FM) which became popular with the discovery of spin-transfer torque (STT)[8,9], in which a spin-polarized charge current was utilized to change the magnetic orientation of a ferromagnet. The major advantage of such an electrical technique is device scalability and reduced power consumption over conventional magnetic field based devices. Consequently, STT was used to build magnetic memories such as STT- magnetoresistive random access memory (MRAM). As illustrated in Fig. 1(a), the basic component of a STT device is the magnetic tunnel junction (MTJ) which consists of an oxide tunnel barrier sandwiched between two ferromagnetic layers. During the device operation, when a charge current is passed through the MTJ, the electrons are spin polarized in one of the FM and are subsequently used to manipulate the magnetic state of the other FM using STT. However, as the charge current tunnels through the insulating oxide barrier, a large write current density can lead to the breakdown of the oxide.[10,11] Thus, substantial efforts are still being devoted to reduce the amount of the current required for manipulation of the magnetization using STT. Moreover, the magnetic state of the MTJ is sensed using the TMR by passing a smaller read current through the MTJ. As a result of the coupling of the read and write current paths, the read current can lead to accident switching of the magnetic states (read disturbance)[11].

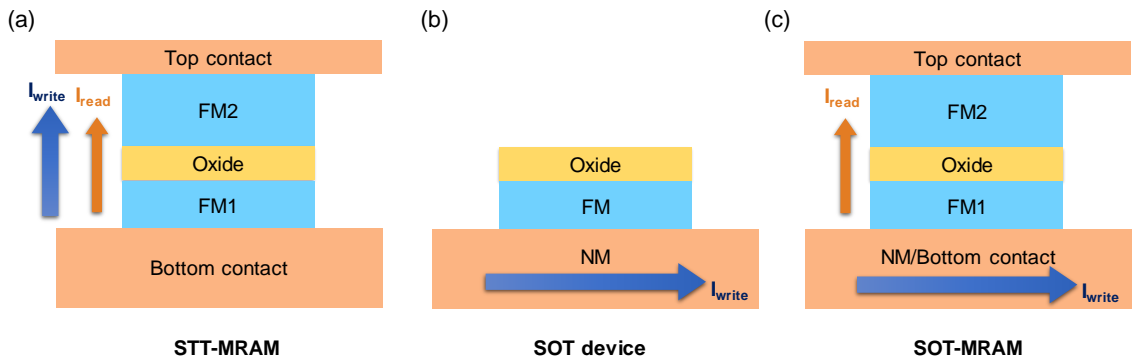

Fig. 1. (a) Schematic of a magnetic tunnel junction in conventional STT-MRAM. The read and write current paths are coupled in the STT-MRAM cell. (b) Schematic of a SOT device illustrating the write current path in the SOT scheme. (c) Schematic of a SOT-MRAM cell utilizing SOT scheme for writing and TMR scheme for readout.

In the recent years, an alternative technique of magnetization manipulation, namely, the spin-orbit torque (SOT)[12-17], was discovered to electrically switch a FM which overcomes the afore-mentioned shortcomings of the STT technique. The technique of SOT utilizes the spin-orbit interaction to generate a spin current and possesses added advantages over STT in terms of reduced power consumption[18] and faster device operation[19,20]. As illustrated in Fig. 1(b), a typical SOT device is composed of a bilayer consisting of a FM and a non-magnetic material (NM) capped by an oxide. When an in-plane charge current injected into the bilayer, a transverse spin current density at the bilayer interface is generated due to the spin-orbit coupling (SOC) effects at the bulk of the NM and/or the interface of NM/FM. This spin accumulation at the interface exerts the torque on the magnetization of the FM and subsequently has been proven to switch the magnetization of the FM, move domain walls inside the FM, and generate oscillations in an effective manner compared with conventional STT. As shown in Figure 1(c), the SOT writing scheme can be combined with the TMR based reading scheme to construct a SOT-MRAM cell. The magnetization of the FM1 is controlled by SOT from the in-plane write current, while the magnetic state is sensed by passing a smaller read current through the MTJ. In contrast to the STT-MRAM scheme illustrated in Fig. 1(a), the read and write current paths are decoupled in the SOT-MRAM in Fig. 1(c), which allows for better design margins. Furthermore, as the large write current does not flow through the MTJ, the SOT-MRAM scheme allows for a better device stability.

Owing to the aforementioned advantages of the SOT scheme, intense research studies are being carried to better understand the physics of SOT which can be further applied to design ultrafast and power efficient spintronic devices. In this review, we will focus on recent advances in SOT studies and highlight some of the technological aspects and challenges in building practical SOT devices. We first discuss the basic physics of SOT including its origins, torque decomposition, and magnetization switching using SOTs. This is followed by a summary of recent studies on novel material choices for the NM and FM for enhancing the SOT efficiency. We then summarize the recent studies on the SOT switching dynamics, which shed light on detailed microscopic switching mechanisms and provide critical parameters of interest for real applications. This is followed by a discussion of the recent techniques to eliminate the requirement of an external magnetic field during SOT switching and their practical difficulties. We conclude by summarizing the review and providing future directions for the SOT research.

## A. Origins of spin-orbit torque

Although the SOT is attributed to arise due to the spin accumulation at the FM/NM interface, the detailed microscopic origins of the spin current generation are under debate and research. The two main SOC phenomena that are attributed to generate the spin accumulation are spin Hall effect (SHE) and interface Rashba-Edelstein effect.

### 1. Spin Hall effect

The phenomenon of SHE exploits the bulk SOC in the NM to convert an unpolarized charge current into a pure spin current. The bulk SOC in the NM arises from either the band structure (intrinsic) and/or addition of high SOC impurities (extrinsic), and gives rise to spin dependent asymmetric scattering of the conduction electrons. This asymmetric scattering leads to deflections of spin-up and spin-down electrons in opposite directions creating a transverse spin current when an unpolarized charge current is injected into the NM. The SHE was theoretically predicted in 1971 by Dyakonov and Perel[21], revived in 1999 by Hirsch[22] and observed using Kerr microscopy[23] in 2004. Figure 2(a) illustrates a spin accumulation generated at the FM/NM interface due to the bulk SHE in a NM. Note that the polarization $\boldsymbol{\sigma}$ of the accumulated spins is orthogonal to directions of both the injected charge current ($\mathbf{J_C}$) as well the generated spin current ($\mathbf{J_S}$). Accordingly, the SHE is represented using the equation, $\mathbf{J_S} = \frac{\hbar}{2e} \theta_{SH} (\mathbf{J_C} \times \boldsymbol{\sigma})$. Here, $\theta_{SH}$ is the intrinsic property of the NM that quantifies its spin current generation efficiency and is called the spin Hall angle of the NM. The magnitude of $\theta_{SH}$ determines the amount of the spin current density a NM can generate for a given charge current density, and the sign of $\theta_{SH}$ determines the direction of the spin accumulation at the NM/FM interface. While the process of SHE generates a pure spin current from an unpolarized charge current; conversely, a charge current can be generated due to a pure spin current in the NM by a reverse process. This reverse process is called the inverse spin Hall effect (ISHE)[24,25] and originates from the same SOC effects as the SHE. For a detailed review of SHE, ISHE and associated mechanisms, the readers are encouraged to refer to reviews by Hoffman[26] and Sinova *et. al*[27].

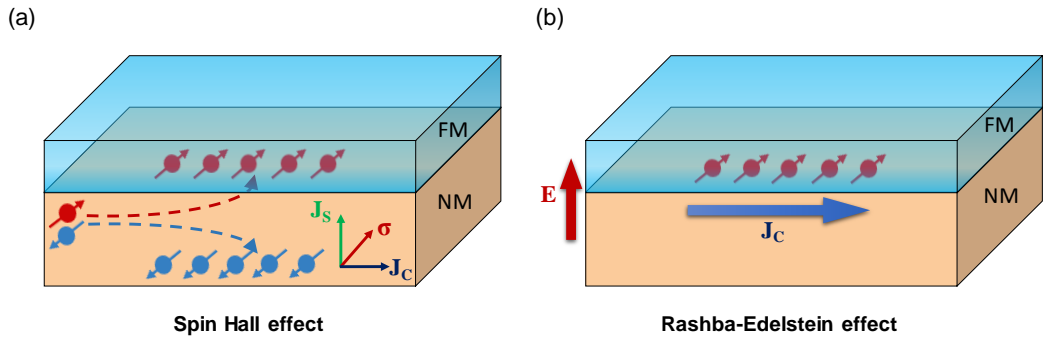

Fig. 2. (a) Illustration of the bulk spin Hall effect in a NM. (b) Illustration of Rashba-Edelstein effect at the FM/NM interface.

## 2. Interface Rashba-Edelstein effect

The Rashba-Edelstein effect (also called inverse spin galvanic effect[28]) originates from an interfacial SOC phenomenon[29,30] that arises in structures with broken inversion symmetry such as NM/FM/Ox in Fig. 1(b), where an internal electric field, $\mathbf{E}$, is generated along the direction of symmetry breaking (see Fig. 2(b)). The conduction electrons with momentum, $\mathbf{p}$, moving near this interface with $\mathbf{E}$, experience an effective magnetic field in the direction $\mathbf{E}\times\mathbf{p}$. This magnetic field couples to spin magnetic moments of the conduction electrons and polarizes their spin magnetic moment along $\mathbf{E}\times\mathbf{p}$. Accordingly, this interfacial SOC (Rashba SOC) is modelled using the Hamiltonian, $H_R = \frac{\alpha_R}{\hbar}(\mathbf{E} \times \mathbf{p}) \cdot \boldsymbol{\sigma}$, where $\alpha_R$ is the Rashba parameter. The Rashba-Edelstein effect to generate spin currents originates from this interfacial SOC which was initially proposed in the context of wurtzite semiconductors[29] and 2 dimensional electron gases[30] with broken inversion symmetry and then extended to NM/FM bilayers over the last decade[13,31,32]. The Rashba effect at the NM/FM interface is illustrated in Fig. 2(b). Similar to the case of SHE and ISHE, the Rashba-Edelstein effect also has its counterpart referred to the inverse Rashba-Edelstein effect (or spin galvanic effect)[28,33,34], where a non-equilibrium spin accumulation generates a charge current due to interfacial SOC. For a detailed review on the Rashba effect and associated spin-orbit torques, the readers can refer to a recent review by Manchon *et al.*[35]

**B. Torque decomposition of SOT**

Irrespective of the underlying origins of the spin accumulation at the NM/FM interface, the SOT exerted on the magnetization ($\mathbf{m}$) of a FM due to this non-equilibrium spin density can be decomposed into two components[36-38], namely the damping-like (or Slonczewski) torque, $\boldsymbol{\tau}_{DL} \sim \mathbf{m} \times (\boldsymbol{\sigma} \times \mathbf{m})$ and the field-like torque, $\boldsymbol{\tau}_{FL} \sim (\boldsymbol{\sigma} \times \mathbf{m})$. From the torque symmetry, $\boldsymbol{\tau}_{DL}$ displays a damping-like behavior that tends to align $\mathbf{m}$ along $\boldsymbol{\sigma}$ and is essentially the Slonczewski spin transfer torque[9] on the magnetization of a FM due to the injected spin current with the polarization $\boldsymbol{\sigma}$. Therefore, intuitively $\boldsymbol{\tau}_{DL}$ was associated to arise from the spin currents generated from bulk SOC effects like SHE and diffusing toward the FM. On the other hand, the effect of $\boldsymbol{\tau}_{FL}$ on $\mathbf{m}$ is similar to that of a magnetic field that makes $\mathbf{m}$ to precess around $\boldsymbol{\sigma}$. Thus, $\boldsymbol{\tau}_{FL}$ was attributed to arise from an exchange interaction from the polarized spins at the NM/FM interface generated from interfacial SOC phenomena like the Rashba-Edelstein effect. However, recent experiments have shown both bulk as well as interfacial SOC effects can give rise to both the components of SOTs. Consequently, it is not straightforward to identify the contribution of individual effects to each torque component of SOTs and requires detailed further experiments, such as a thickness dependence of the FM and NM, insertion spacer layer dependence between the NM and FM, and so on. Apart from the fundamental physical understanding, it is important for applications to identify bulk vs. interfacial contributions in different material systems to determine the appropriate material parameters that need to be tuned for engineering the torque components of SOTs. In experiments, the quantities that are characterized are the effective magnetic fields[39-42] due to these torque components, $\mathbf{B}_{DL}$ and $\mathbf{B}_{FL}$, given by the equation $\boldsymbol{\tau}_{DL,FL} = \mathbf{m} \times \mathbf{B}_{DL,FL}$ as illustrated in Fig. 3(a). There are many different techniques to evaluate the magnitude of the SOT effective fields. The most commonly used methods are harmonic Hall voltage measurements[39-44], spin torque ferromagnetic resonance (ST-FMR) measurements[17,45,46], and magneto-optic Kerr effect[47-49].

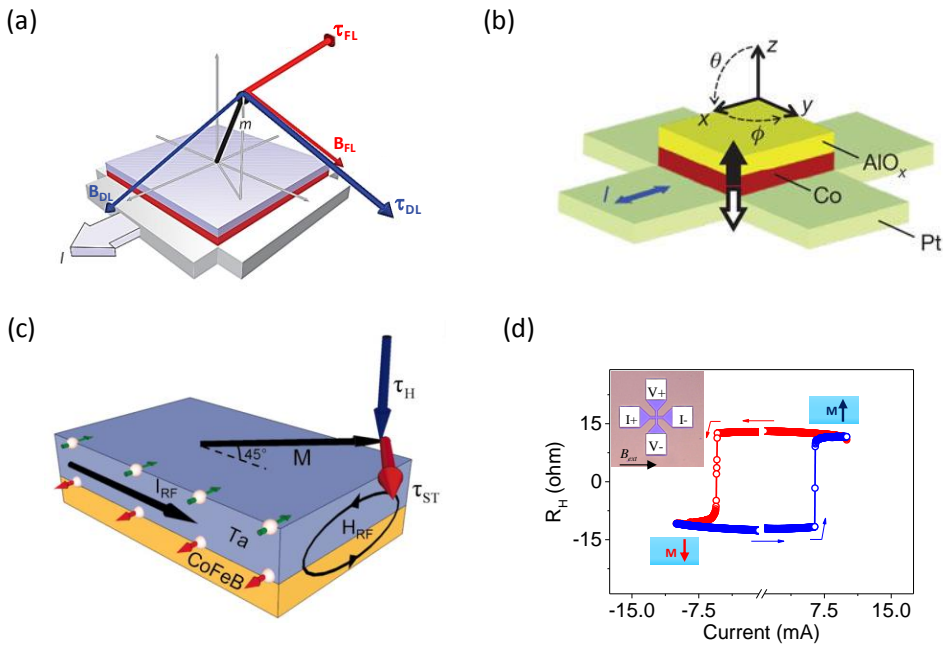

Fig. 3. (a) Illustration of the two spin-orbit torque components and their respective effective magnetic fields. Adapted by permission from Springer Customer Service Centre GmbH: Springer

Nature, Nature Nanotechnology **8**, 587, K. Garello *et al.*, Copyright 2013. (b) Schematic illustration of SOT induced magnetic switching in a perpendicularly magnetized Co dot on top of a Pt channel. Reprinted by permission from Springer Customer Service Centre GmbH: Springer Nature, Nature **476**, 189, I. M. Miron *et al.*, Copyright 2011. (c) Illustration of SOT on an in-plane CoFeB adjacent to the heavy metal Ta. From L. Liu *et al.*, Science **336**, 555 (2012). Reprinted with permission from AAAS. (d) An example SOT induced magnetization switching in a PMA system sensed from the anomalous Hall resistance. The inset shows measurement configuration for Hall resistance measurements.

## C.  SOT induced magnetization switching

The most promising application of SOT is to switch the magnetization state of the FM in a deterministic manner and much faster in comparison with conventional STT scheme.[19,50] Although the SOT can switch FMs with both perpendicular magnetic anisotropy[14] (PMA, Fig. 3(b)) and in-plane anisotropy[15] (IPA, Fig. 3(c)), the current MRAM implementations prefer PMA over IPA due to better scalability[51]. Thus, we will predominantly focus on the SOT switching in the PMA structures in this review. Figure 3(d) shows a representative magnetization switching loop driven by in-plane currents in a Ta/CoFeB/MgO system with PMA. Depending on the polarity of the current, the magnetization state (sensed by the anomalous Hall resistance) of the FM can be either up or down. However, in order to switch PMA using SOTs with the spins having in-plane polarization, it is necessary to apply an external magnetic field[14-16,52,53] ($B_{ext}$ in the inset of Fig. 3(d)) to break the symmetry in the system unlike the STT scheme. This requirement of an external magnetic field is one of the major bottlenecks that is hindering the use of SOTs in practical applications. In section IV, we will discuss the different techniques that have been proposed to mitigate this issue of external magnetic field as well as the feasibility of these techniques.

For quantitative comparison of the SOT switching in different systems, it is useful to define the SOT switching efficiency. Early studies assumed that the SOT switching process follows a macrospin behavior where the magnetization of the FM coherently switches in a uniform manner. In such a scenario for the PMA magnet, the in-plane spins compete with the anisotropy ($H_K$)[16,19,52] of the FM during the switching. Thus, the SOT switching efficiency[52,54-58] can be defined as $\eta = H_K / J_C$ or $\eta = (H_K - \sqrt{2}H_{ext})/J_C$, where $H_{ext}$ is the magnitude of the external assist field and $J_C$ is the critical switching current density. However, as we will discuss in section III, in large samples (> ~ 50 nm), the switching does not follow the above described macrospin behavior, but instead follows domain nucleation and expansion model. Under this condition, a better parameter to define the efficiency is using the depinning field, $H_P$. In the domain mediated switching, the SOT switching efficiency[59,60] can be described as $\eta = H_P / J_C$.

## II. Material choices for SOT devices

### A. Metallic non-magnetic layers

Traditionally, the NM underlayer for the SOT studies is a heavy metal owing to a large SOC strength. The mostly widely used NM heavy metals for SOT devices include Pt[14,17], Ta[15,39], W[61,62], Hf[55,63]. Among these heavy metals, W in $\beta$-phase[61] shows the largest spin Hall angle of −0.3. Further, the NM thickness dependence studies[39,55,62] for cases of Ta and Hf underlayers have suggested there is a competition between the interface and the bulk SOC effects in the contribution to SOTs. The interface contributions dominate in the thinner NM regime and the bulk effects dominate in the thicker NM regime. Moreover, due to the opposite nature of these contributions in the Hf and Ta systems, as the NM thickness increases, there is a sign change in the observed SOTs (at ~0.6 nm for Ta and at ~2 nm for Hf). Apart from single heavy metals, bilayer of heavy metals, such as Pt/Ta[64,65], Pt/W[65], Pt/Hf[66], W/Hf[67] are also explored. It was found that a thin Hf layer insertion[66,67] between the NM and FM can improve the interfacial properties, such as enhancement of PMA (surface anisotropy energy increases by two orders) and reduction of Gilbert damping (by a factor of 2), while maintaining the SOT efficiency. Furthermore, studies have found that the SOT strength can be enhanced by sandwiching the FM between two NMs[57,68,69] or modulating the spin absorption by using a proper capping layer[70] such as Ru.

Apart from heavy metals, metal alloys have also been explored. In particular, it has been shown that addition of impurities in a light metal, such as Cu, can result in a large spin current generation efficiency via. the extrinsic spin Hall effect[71,72]. Furthermore, as Cu is the most common metallization element in CMOS, Cu alloys-based spintronic devices offer ease of integration into the existing Si fabrication technology. Accordingly, the spin Hall effect and associated SOTs have been explored in Cu based alloys such as CuBi[73,74], CuIr[73,75-77], CuPb[73], CuPt[78], CuAu[79-81]. Figure 4(a) shows the non-local spin valve resistance ($R_S$)[73-75,82] for the case of $Cu_{99.5}Bi_{0.5}$. The change in the $R_S$ is reduced for the case of $Cu_{99.5}Bi_{0.5}$ compared to case without any metal, indicating a strong spin Hall effect in these alloys. Notably, $Cu_{99.5}Bi_{0.5}$ shows a large $\theta_{SH}$ of −0.24, which is larger than the measured $\theta_{SH}$ in heavy metals Pt and Ta. Figure 4(b) shows the spin Hall angle extracted for an another Cu based alloy CuPt as a function of Pt concentration[78]. It is observed that the spin Hall angle increases as the Pt concentration increases and even ~28 % Pt in CuPt can give rise to a spin Hall angle closer to that of pure Pt. For a detailed review of extrinsic SHE in such alloyed systems, the readers can refer to a recent review by Niimi *et al.*[82] Apart from Cu-based alloys, studies have found that alloying can enhance the spin current generation efficiency in heavy metals as well. Accordingly, alloys such as AuW[83,84], PtHf[85,86], PtAl[85], and AuTa[84,87] have been studied in the recent works. Interestingly, for the case of AuTa[84], it was found that 10% Ta can give rise to a larger $\theta_{SH}$ of 0.5 compared to the $\theta_{SH}$ of ~0.3 observed in W. Moreover, the $Au_{90}Ta_{10}$ shows a lower resistivity of $85\ \mu\Omega \cdot cm$ compared to W ($\sim 170-260\ \mu\Omega \cdot cm$)[61] and thus can have reduced power consumption as well.

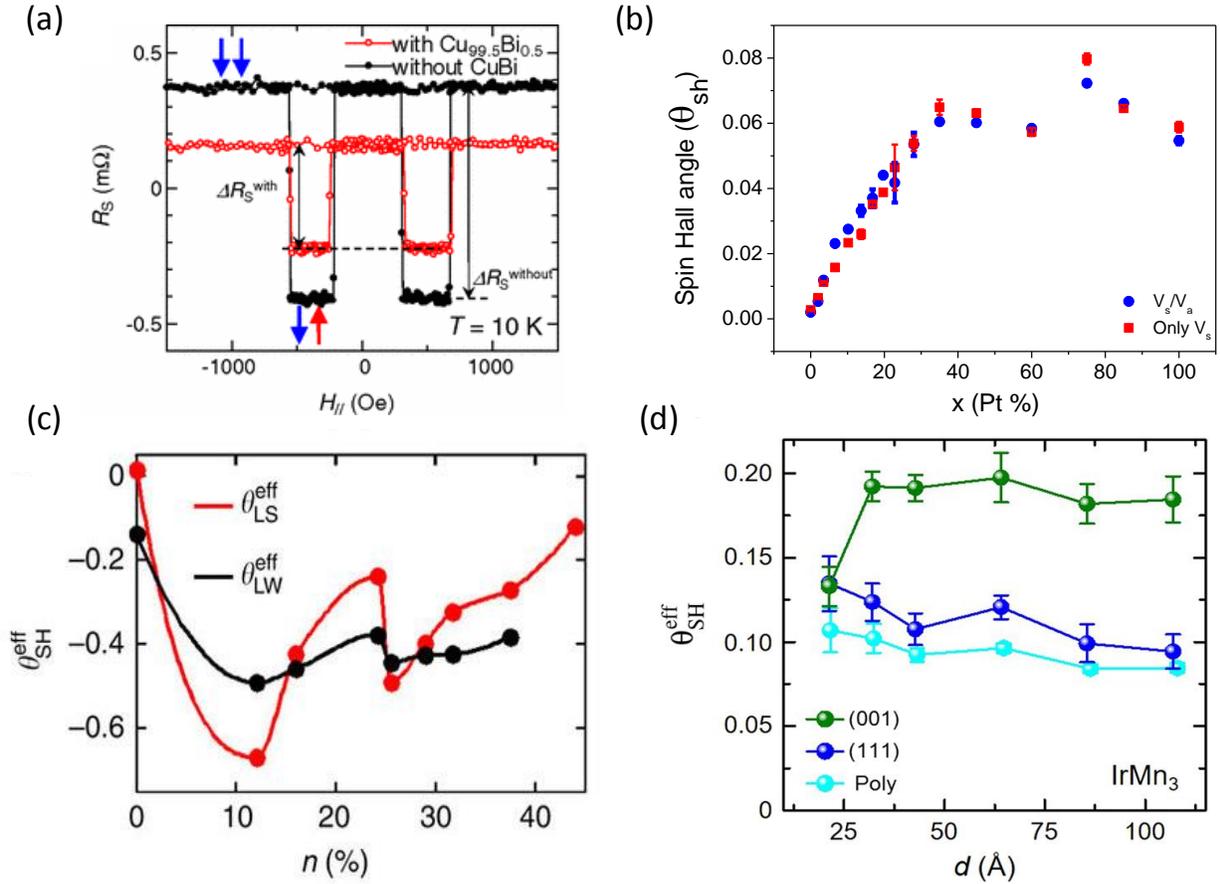

Fig. 4. (a) Nonlocal spin valve signals measured at 10 K for $Cu_{99.5}Bi_{0.5}$ and for without any metal. The change in the $R_S$ is reduced with an insertion of $Cu_{99.5}Bi_{0.5}$ indicating a strong ISHE. Reprinted figure with permission from Y. Niimi *et al.*, Phys. Rev. Lett. **109**, 156602 (2012). Copyright (2012) by the American Physical Society. (b) Spin Hall angle measured as a function of Pt concentration in a CuPt system. Reprinted figure with permission from R. Ramaswamy *et al.*, Phys. Rev. Appl.

**8**, 024034 (2017). Copyright (2017) by the American Physical Society. (c) Effective spin Hall angle extracted from ST-FMR using line shape analysis ($\theta_{LS}^{eff}$) and line width analysis ($\theta_{LW}^{eff}$) for W(O) as a function of oxygen concentration (n). Reused from K.-U. Demasius *et al.*, Nat. Commun. **7**, 10644 (2016), which is licensed under a Creative Commons Attribution 4.0 International License. (d) Effective spin Hall angle as a function of IrMn$_3$ thickness for different crystal orientations. Reused from W. Zhang *et al.*, Sci. Adv. **2**, e1600759 (2016), which is licensed under Creative Commons Attribution-NonCommercial license.

Recent studies have also found that the spin current generation efficiency can be enhanced in a metal by oxygen incorporation[56]. For example, oxygen incorporation has been shown to modify the grain structure and substantially enhance the spin current generation efficiency in W(O)[88]. Moreover, as illustrated in Fig. 4(c), a large spin Hall angle of −0.5 was obtained in W(O) and the spin Hall angle showed a weak oxygen concentration dependence upto 44 % oxygen in W(O). Similarly, oxygen incorporation via. natural oxidation was utilized to enhance the spin Hall angle of the light metal Cu to ~0.08 which is comparable to that of pure Pt.[89] Recent studies have also found sizable SOTs in metallic antiferromagnets[90-95], such as PtMn and IrMn. The SOT efficiency from epitaxially grown antiferromagnets showed an anisotropic behavior with respect to their crystal orientations[94,95]. An example of such an anisotropy is illustrated in Fig. 4(d) in the case of IrMn$_3$ system, where the measured effective spin Hall angle shows different magnitudes depending on the crystal orientation and structure. Further, these AFM materials can also act as a source to provide the assistive magnetic field via. exchange bias for realizing external field free magnetization switching using SOTs as we will discuss in Sec. IV B.

## B. Exotic non-magnetic layers

While heavy metals and their alloys have been a standard choice as the NM for SOT generation, exotic materials such as topological insulators (TIs)[45,96-105], transition-metal dichalcogenides (TMDs)[106-110] or even a two dimensional electron gases (2DEGs)[111-115] have shown to possess a large spin current generation efficiency.

TIs are quantum materials that possess a bandgap like an insulator but have topologically protected conducting edge states. These conducting states are called topological surfaces states (TSS) which possess spin-momentum locking which causes the electrons moving on TSS to have their spin polarization locked in the orthogonal direction to their motion. Consequently, TIs have a very high spin current generation efficiency and are actively pursued for spintronic applications. In literature, many works[45,96-105] have studied spin current generation and associated SOTs in different TI materials. The estimated values of spin Hall angle from TIs are 2−3 orders of magnitude larger than that in conventional heavy metals. Figures 5(a) and 5(b) show the temperature dependence of the damping-like and field-like torque components of SOTs, respectively, in a $Bi_2Se_3$/Py system[97]. While the appearance of field-like torque in $Bi_2Se_3$/Py is expected due to the interfacial nature of the generated spins from TSS, the appearance of damping-like torque indicates a possible bulk SHE in this system. However, the observed $\tau_{DL}$ increases substantially as the temperature decreases (Fig. 5(a)), which is not expected from bulk SHE mechanism. Therefore, $\tau_{DL}$ was attributed arise from the spin transfer of TSS induced spins[45,97]. Furthermore, it is noted that the Rashba effect in the 2DEG of $Bi_2Se_3$ could also induce a spin accumulation. However, the experimentally observed sign of the spin polarization was opposite to that from the Rashba effect and was thus concluded to arise from TSS. Subsequently, in a later studies, this large spin current generation efficiency from TIs was used to demonstrate highly efficient SOT induced magnetization switching[96,98,116,117].

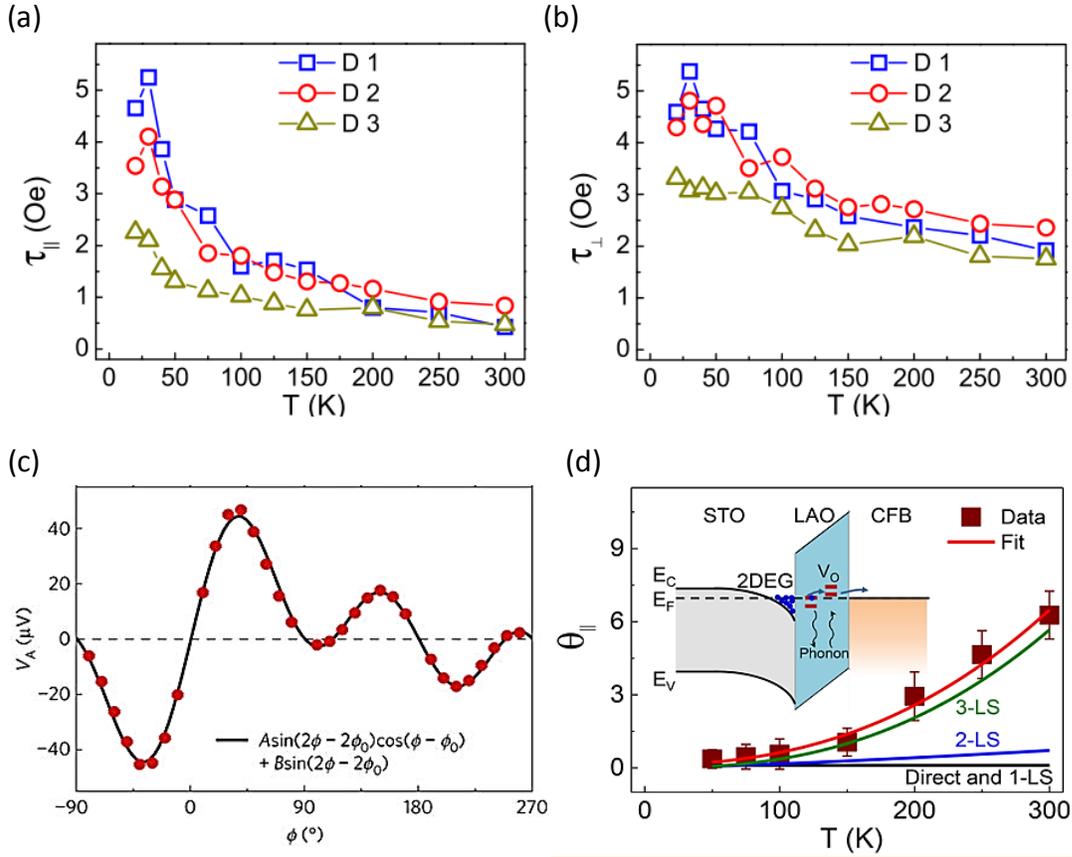

Fig. 5. Temperature dependence of in-plane (a) and out-of-plane (b) SOT (equivalent to damping-like and field-like torque, respectively) in a $Bi_2Se_3$/CoFeB (5 nm) film estimated using ST-FMR technique. Reprinted figures with permission from Y. Wang *et al.*, Phys. Rev. Lett. **114**, 257202 (2015). Copyright (2015) by the American Physical Society. (c) Angular dependence of antisymmetric component of ST-FMR signal for $WTe_2$ (5.5 nm)/Py (6 nm) system fitted with two terms. The term B arises from an out-of-plane damping torque that arises crystal symmetry of $WTe_2$. Adapted by permission from Springer Customer Service Centre GmbH: Springer Nature, Nature Physics **13**, 300, D. MacNeill *et al.*, Copyright 2016. (d) Temperature dependence of spin Hall angle in STO/LAO/CoFeB system. The inset shows a schematic of the spin-polarized electron

inelastic tunneling process via. the localized states such as oxygen vacancies. Reprinted with permission from Y. Wang *et al.*, Nano Lett. **17**, 7659 (2017). Copyright (2017) American Chemical Society.

Very recently, 2D TMDs are receiving immense research attention as their thickness can be reduced to as low as a monolayer. In $MoS_2$/Py, the observed SOTs were attributed to arise from the interface[107]. Another interesting result[109,110] was observed in a layered TMD $WTe_2$, where SOTs were controlled using crystal symmetry. Moreover, it was shown that $WTe_2$ can exert an out-of-plane (OOP) damping-like torque (see Fig. 5(c)) which can be utilized to switch a magnet with OOP anisotropy without any assistive magnetic field. Recently, SOTs were also observed in a 2DEG formed at the interface of $SrTiO_3$ (STO) and $LaAlO_3$ (LAO),[111-114] which is known to possess a strong Rashba SOC. A large spin Hall angle of ~6.3 at room temperature was estimated in the STO/LAO/CoFeB system[113] and it was concluded from temperature dependent analyses in Fig. 5(d) that inelastic tunneling via localized states, such as oxygen vacancies in the LAO band gap, served as a medium for spin transmission in the insulator LAO. Moreover, STO/LAO 2DEG channel is also reported to have a long spin diffusion length[118,119] of 300 nm. Hence, these oxide materials can find use in future oxide based spintronic devices.

While the above exotic materials possess an extremely large spin current generation efficiency, one of the practical challenges is to grow a FM with PMA on top of these materials. A recent research study has shown that a PMA ferromagnet can be grown on top of a TI by proper material choices[116,117]. Another practical issue in terms of practical integration of these exotic materials into the existing Si platform is material compatibility. For example, research is needed on the high temperature stability of these exotic materials as the industrial Si processes involve high back end annealing temperatures.

## C. Ferromagnets and ferromagnetic multilayers

So far, we discussed the different NMs that have been explored for SOT generation. In terms of the FMs, the preferred FM for SOT applications is the ultrathin CoFeB layer as it can be interfaced with an MgO layer for strong interfacial PMA and a large TMR. Apart from CoFeB, the other commonly studied single layer FMs in SOT studies include permalloy (NiFe), pure Co, and CoFe. One of the challenges with single layer FMs is that for inducing OOP magnetization, interfacial PMA is generally used which requires the FM layers to be ultrathin, such as, less than 1.4 nm typically. Such thickness requirements restrict lateral shrinkage of the magnet, for certain applications such as MRAMs, where there is a requirement of minimum thermal stability factor which is proportional to the magnetic volume.

Alternatively, ferromagnetic multilayers, such as Co/Ni[20,69,70,120], Co/Pd[121], and Co/Pt[122] have been explored for studying SOTs as they can be grown very thick while retaining their PMA. However, for the case of FMs, the strength of SOTs scales inversely with the thickness of a FM. Hence, there is a tradeoff between the thermal stability and SOT efficiency. Nevertheless, it has been shown in an early study[121] that the PMA FM multilayers such as Co/Pd can be grown very thick (~20 nm, see Fig. 6(a)) and still possess very large SOT effective fields ( $|\mathbf{B}_{DL}| = 1170$ Oe $/10^8$ A cm$^{-2}$ and $|\mathbf{B}_{FL}| = 5025$ Oe $/10^8$ A cm$^{-2}$, which are 5–10 times larger than the CoFeB case[39,40]), indicating a possible bulk origin of SOTs in FM multilayers. In addition, synthetic antiferromagnets (SAFs) based on the FM multilayers ([Co/Ni/Co]/Ru/[Co/Ni/Co])[20] have been used for racetrack memory applications, as SOTs in combination with exchange torques from the RKKY interaction have been shown to move the domain walls very efficiently. Figure 6(b) shows that the SOT driven domain wall motion can reach large velocities of 750 ms$^{-1}$ in a Co/Ni/Co based SAF with a Pt underlayer. Recently, an anomalous switching behavior[123] was observed in SAF structures, where the sign of effective SOTs can be tuned by the strength of the external magnetic field. This is illustrated in Fig. 6(c) where the different magnetization states in a [Co/Pt]$_n$ based SAF can be achieved by changing the magnitude of the assist field for a fixed polarity of current and assistive magnetic field.

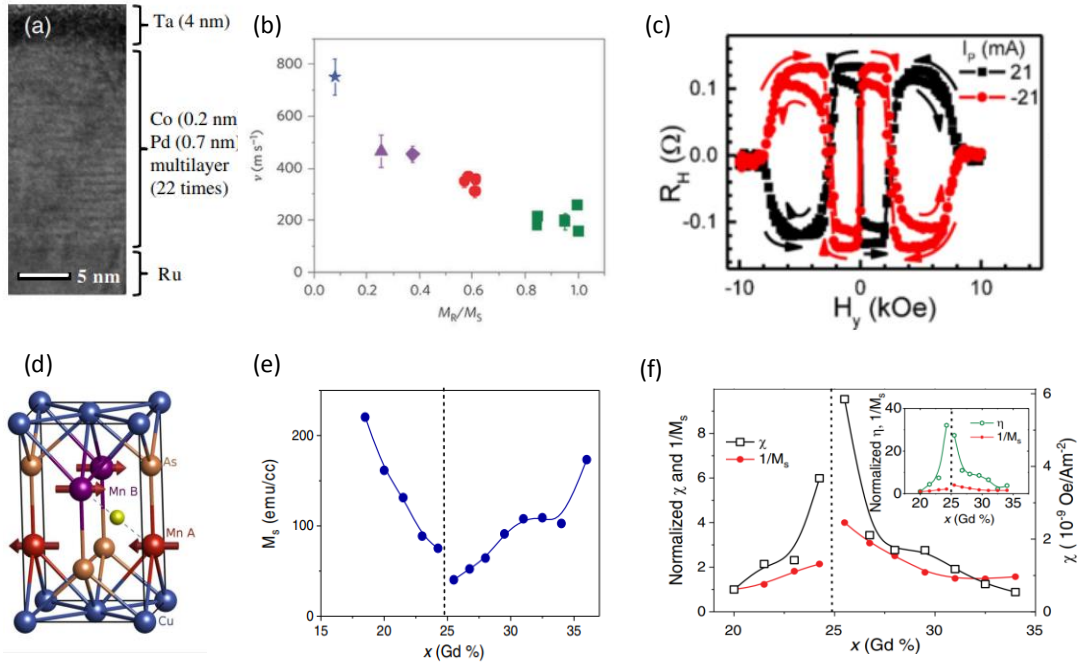

Fig. 6. (a) Transmission electron microscopy image of [Co/Pd]$_{22}$ multilayer with PMA. Reprinted figure with permission from M. Jamali *et al.*, Phys. Rev. Lett. **111**, 246602 (2013). Copyright (2013) by the American Physical Society. (b) SOT driven domain wall velocity as a function of the ratio of remanent magnetization (M$_R$) with respect to the M$_S$ in a [Co/Ni/Co] based SAF. The smaller the value M$_R$/M$_S$, the larger the antiferromagnetic strength and hence the faster domain wall velocity. Reprinted by permission from Springer Customer Service Centre GmbH: Springer Nature, Nature Nanotechnology **10**, 221, S.-H. Yang *et al.*, Copyright 2015. (c) The Hall resistance with a current amplitude of ±21 mA as a function of in-plane assist field in a [Co/Pt]$_n$ based SAF.

Reprinted figure with permission from C. Bi *et al.*, Phys. Rev. B **95**, 104434 (2017). Copyright (2017) by the American Physical Society. (d) Crystal structure of the antiferromagnet CuMnAs. Due to local crystal inversion asymmetry, the two Mn sites A and B generate opposite spin polarizations, which have been used to switch between the states of the CuMnAs. From P. Wadley *et al.*, Science **351**, 587 (2016). Reprinted with permission from AAAS. (e) The $M_S$ as a function of the Gd concentration in a CoGd ferrimagnet. Reprinted figure with permission from R. Mishra *et al.*, Phys. Rev. Lett. **118**, 167201 (2017). Copyright (2017) by the American Physical Society. (f) The SOT switching efficiency (defined in Sec. I C) and $1/M_S$ as function of Gd concentration, normalized with respect to a Gd concentration of 20 %. The macrospin SOT switching efficiency is plotted in the inset. Both the switching efficiencies scale at a more rapid rate compared to $1/M_S$ near the magnetic compensation due to the exchange torque. Reprinted figure with permission from R. Mishra *et al.*, Phys. Rev. Lett. **118**, 167201 (2017). Copyright (2017) by the American Physical Society.

### D. Ferrimagnets and antiferromagnets

Apart from the conventional FMs that have a positive exchange coupling, recent SOT experiments have utilized magnets with negative exchange coupling, such as antiferromagnets[124,125] and ferrimagnets[60,126-131]. Figure 6(d) shows the crystal structure of an antiferromagnet CuMnAs[124], where the local non-equilibirium spin density from the two Mn sites has been used to successfully switch between the two states of the AFM. In addition to pure antiferromagnets, SOT studies have also used ferrimagnets, in particular rare earth-transition metal (RE-TM) ferrimagnets, for the magnetic layers. By tuning the composition of individual constituents of the RE-TM, the magnetization of the RE-TM ferrimagnet can be varied. As illustrated in Fig. 6(e), for the case of $Co_{1-x}Gd_x$, near the magnetization compensation ($x \approx 25\%$), the value of saturation magnetization ($M_S$) is reduced drastically[60]. Close to the compensation point, the negative exchange coupling strength is also enhanced. The composition dependence of SOTs in the ferrimagnet CoGd has found that the strength of the SOT effective field and thus, the SOT switching efficiency (Fig. 6(f)) can be enhanced dramatically near their magnetic compensation. The SOT effective field increases ~9 times and the switching efficiency increases ~6 times near compensation compared to the uncompensated case. Although SOTs scale inversely with the $M_S$, which decreases as we approach compensation, it was observed that the SOTs scale disproportionately and increase much more than the amount of decrease in the saturation magnetization. This significant increase was attributed to the additional torque in ferrimagnet due to the negative exchange coupling which can dramatically enhance the SOT[20,60]. These ferrimagnets have bulk perpendicular anisotropy, therefore they can be grown very thick for a larger thermal stability.[132] However, as discussed earlier, the SOTs also scale inversely with the thickness of magnets. A recent ferrimagnet thickness dependence of SOT revealed that SOTs can switch even a 30 nm thick GdFeCo[127]. Other than Gd based ferrimagnets, Tb based ferrimagnets, such as CoTb were also explored[126,128,129,133]. A composition dependent SOT study[126] in a Ta/CoTb system revealed that the SOT efficiency increases near magnetic compensation. They also find that the Dzyaloshinskii-Moriya energy increases as the Tb concentration increases. Temperature dependent SOT studies in GdFeCo[130] and CoTb[128] revealed that the damping-like SOT showed a strong temperature dependence and increases near the magnetic compensation temperature, while the field-like SOT does not change much with temperature.

## III. SOT switching dynamics

As we discussed in Sec. I C, the initial theories to explain the current induced SOT switching phenomena proposed a macrospin coherent rotation model[16,52,134-136] in which the magnetization of a FM is uniform throughout the switching process. Although the macrospin models could qualitatively explain the SOT switching phenomena including the requirement of assist fields[16,53] for deterministic switching, the experimentally observed switching current densities[19] were substantially smaller than the predicted switching current densities from the macrospin models. This observation hinted that the SOT switching process proceeds via. domain nucleation and expansion[16,59,69,137-142] for the case of large magnets (of the order of few hundreds nm and above). Further, the initial simulations[52,134,142] assumed negligible or moderate role of $\tau_{FL}$ in the deterministic switching process and later theories[54,135,136,143] suggested $\tau_{FL}$ plays a dominant role in the switching dynamics. Consequently, recent works studies the SOT switching dynamics studies[138,144-147] with short electrical pulses using enhanced spatial and time resolution to understand the microscopic switching processes and shed light on the exact roles of $\tau_{DL}$ and $\tau_{FL}$ in the switching process. In addition to enhanced understanding of SOT switching physics, the study of switching dynamics provides a better quantitative picture that is necessary for reliable operation of SOT devices. In the following sections, we summarize the key studies of SOT switching dynamics in spatial and time domain, and highlight their important findings.

### A. Short pulse current injection

The early works[14-16] on the current induced magnetization switching using SOTs operated in the quasi-static regime where the applied current pulse duration was very long. In contrast, the magnetization dynamics operate in the order of a few ns and below. Therefore, the first step to study the SOT induced magnetization dynamics is to reduce the injected current pulse duration to ns regime. In the studies of magnetic switching dynamics, the stochasticity of the switching process becomes highly relevant and hence the results are usually expressed in terms of switching probabilities. This is illustrated in Fig. 7(a) where the SOT switching probability is plotted as a function of pulse duration ($\tau_p$) for different current magnitudes for a Pt (3 nm)/Co (0.6 nm)/AlOx nanodot[19]. Here, the switching probability was obtained by averaging the difference in the Hall resistance before and after an application of the pulse over 100 trials. It is observed that as the current magnitude increases, the pulse duration required to achieve 100% switching probability decreases. The authors subsequently extracted the critical switching current ($I_C$ - current required for 90% switching probability) for different pulse durations as shown in Fig. 7(b).

From Fig. 7(b), the following key findings were identified. First, there are two time-scale regimes (similar to conventional STT[148,149]): (i) For pulse duration < 10 ns, where the critical current increases significantly with decreasing the pulse duration, the switching dynamics are dominated by the current induced torques. (ii) For pulse duration > 10 ns, where the critical current shows a weak dependence on the pulse duration, the switching dynamics are dominated by stochastic thermal fluctuations as the pulse duration is long enough for thermal effects to play a role. Second, the data in Fig. 7(b) for short time scales (< 1 ns) were fitted to the equation[148,149], $I_C = I_{C0} + \frac{q}{\tau_p}$ to obtain an intrinsic critical current density, $I_{C0}$. Here, $q$ is interpreted as the efficiency of angular momentum transfer into the system. The estimated $I_{C0}$ (0.58 mA) was smaller than the values predicted from macrospin models (2.05 mA)[52,134]. This suggests that the switching in ~90 nm Pt/Co nanodot proceeds via. domain nucleation and propagation further confirming the previous observations in the quasi-static experiments[16,59]. Finally, the authors find that the incubation time[148,149] of the SOT switching is negligibly small (~ $10^{-20 \pm 2}$ s).

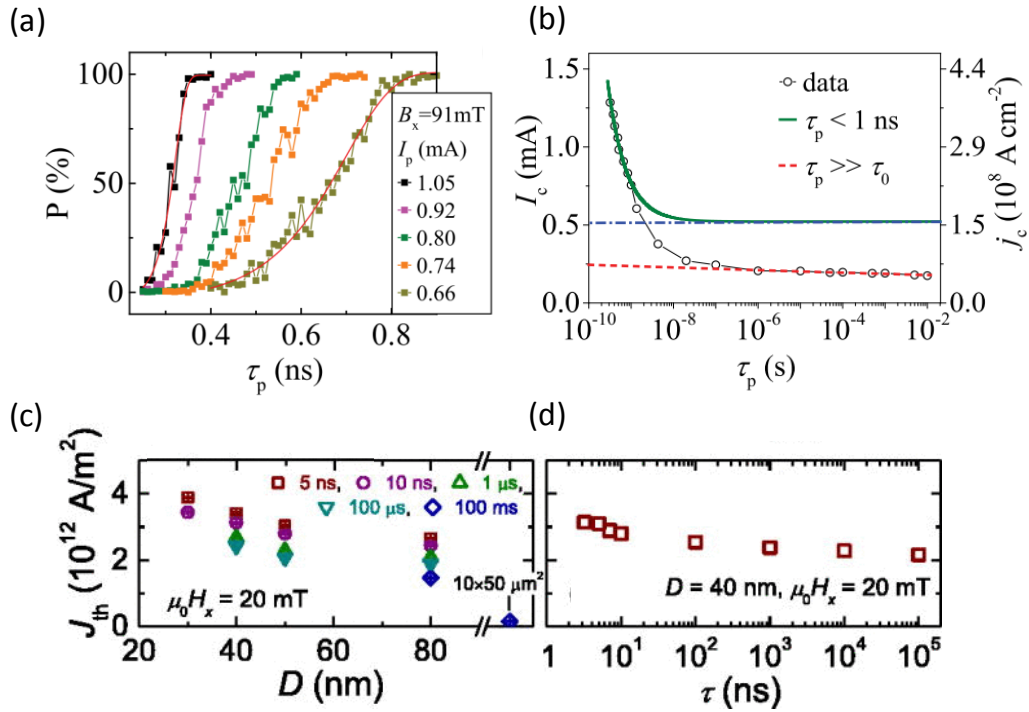

Fig. 7. (a) Switching probability for a Pt (3)/Co (0.6)/AlOx nanodot (90 nm lateral size) as a function of applied pulse duration. Reprinted from K. Garello *et al.*, Appl. Phys. Lett. **105**, 212402 (2014), with the permission of AIP Publishing. (b) Critical SOT switching current (left y-axis) and current density (right y-axis) for a Pt (3)/Co (0.6)/AlOx nanodot (95 nm lateral size) as a function of applied pulse duration. Reprinted from K. Garello *et al.*, Appl. Phys. Lett. **105**, 212402 (2014), with the permission of AIP Publishing. (c) Device diameter (D) dependence of critical SOT switching current density in Ta (5)/Co$_{18.75}$Fe$_{56.25}$B$_{25}$ (1.2)/MgO (1.5) nanodots. Reprinted from C. Zhang *et al.*, Appl. Phys. Lett. **107**, 012401 (2015), with the permission of AIP Publishing. (d) Critical SOT switching current density as a function of pulse duration for a Ta (5)/Co$_{18.75}$Fe$_{56.25}$B$_{25}$ (1.2)/MgO (1.5) device with 40 nm diameter. Numbers in parentheses are in nm. Reprinted from C. Zhang et al., Appl. Phys. Lett. **107**, 012401 (2015), with the permission of AIP Publishing.

In general, as the lateral dimensions of the device decrease, it is understood that the switching behavior transitions from the incoherent to coherent macrospin regime[16]. However, as illustrated in Fig. 7(b), even for a lateral dimension of 90 nm, the SOT switching does not follow coherent single domain rotation. Thus, in order to elucidate the validity of the single domain picture, the device sizes need to be further shrunk. Subsequently, a later study[138] explored a device size dependence of SOT switching in Ta/Co$_{18.75}$Fe$_{56.25}$B$_{25}$/MgO for different pulse durations in Fig. 7(c). As the device diameter reduces from 80 to 30 nm, the critical current density for a given pulse duration increases, indicating a transition from an incoherent to coherent regime. This transition to a coherent regime is also evident in Fig. 7(d) which shows the critical current density for 40 nm diameter nanodot for different pulse durations. It is observed that, for pulse durations < 10 ns, the critical current density only slightly increases as the pulse duration decreases as predicted by a macrospin-based model[134], in contrast to 90 nm nanodot shown in Fig. 7(b). The authors also find that although there is qualititative agreement with the macrospin model, the calculated values of spin Hall angle and effective anisotropy for the 40 nm nanodot did not quantitatively agree with the macrospin model. This was because the macrospin models considered the role of $\boldsymbol{\tau}_{DL}$ only and the role of $\boldsymbol{\tau}_{FL}$ is also important for quantitative agreement. Furthermore, in an another short pulse switching measurement study[18], it was found that the Oersted field from the NM channel can also play a significant role for SOT switching in a FM with IPA and can speed up the switching process.

**B. Spatial and time resolved measurements**

As discussed in the short pulse measurements above, for large magnetic structures, the SOT switching proceeds via. domain nucleation and propagation. This invalidity of macrospin approximation for large magnetic dots was also observed in a few quasi-static experiments.

Alternatively, micromagnetic simulations to explain microscopic details of the SOT switching process found the critical role of Dzyaloshinskii-Moriya interaction[150,151] (DMI) in the switching process. The DMI field can result in stabilization of chiral Néel domain walls in the ultrathin magnet during the switching process and the role of the assistive magnetic field in deterministic switching was attributed to break the chirality enforced by DMI. This is illustrated in Fig. 8(a) in reversed magnetic domain (initial state – blue, reversed state – red) in which the presence of DMI leads to stabilization of chiral Néel walls.[59] Without an assistive field ($H_X = 0$), the effective field indicated as $H_{SH,z}$ in Fig. 8(a) due to $\tau_{DL}$ on the center of the domain wall is out-of-plane $(\sigma \times m)$ and it is observed the this effective field is in opposite directions on the either side of the reversed domain. As a result, the SOT will be effective in displacing the domain rather than expansion of the reversed domain. For the case in Fig. 8(b), with an assistive magnetic field stronger than the internal DMI field, the center of the domain wall points in the direction of the assistive field. Thus, for a strong assistive magnetic field as shown in Fig. 8(b), the $H_{SH,z}$ on all sides of the domain points to the same direction leading to the expansion of the reversed domain for switching.

While the above picture explains the domain expansion due to SOT during the switching process, the switching process can either be initiated from multiple random nucleation sites or from a single nucleation site at the edge of the device. To better understand the microscopic processes spatially resolved techniques included scanning magneto optic Kerr microscopy (MOKE)[69,141,152,153] and X-ray magnetic circular dichroism[144] were employed. Figure 8(c) illustrates a recent study[144] (25 nm spatial resolution and 100 ps temporal resolution) which shows a direct observation of microscopic picture of SOT switching dynamics in Pt/Co/AlO$_X$ dots (500 nm diameter). It shows the spatial evolution of the switching process at every 100 ps for 2 ns long current pulse with opposite directions of the applied field and current directions. For each case, the red dots indicate the nucleation point and green arrows indicate the direction of domain wall motion during the switching. It is noted that the nucleation point in this case was always on the left or the right edge as determined by the combined actions of DMI, assist field and $\tau_{DL}$. The top-bottom asymmetry in the nucleation point was attributed to $\tau_{FL}$. The authors find that $\tau_{FL}$ enhances the switching efficiency and is also an important factor in determining the domain wall tilt angle during the switching process.

While this above study finds that the domain nucleation takes place at the edge, a recent experiment of time resolved scanning MOKE of SOT switching in Pt/Co finds that by considering the role of sample heating during current applications, the switching may start with the random nucleation of small magnetic bubbles.[147] Further, their micromagnetic simulations hint that the combined role of DMI and $\tau_{FL}$ may also result in long switching times that were observed in their experiments. Furthermore, a recent size dependence study[154] of SOT switching in a W/CoFeB/MgO structure also finds that the nucleation in their study is random for larger device diameters, such as 700 nm and above.

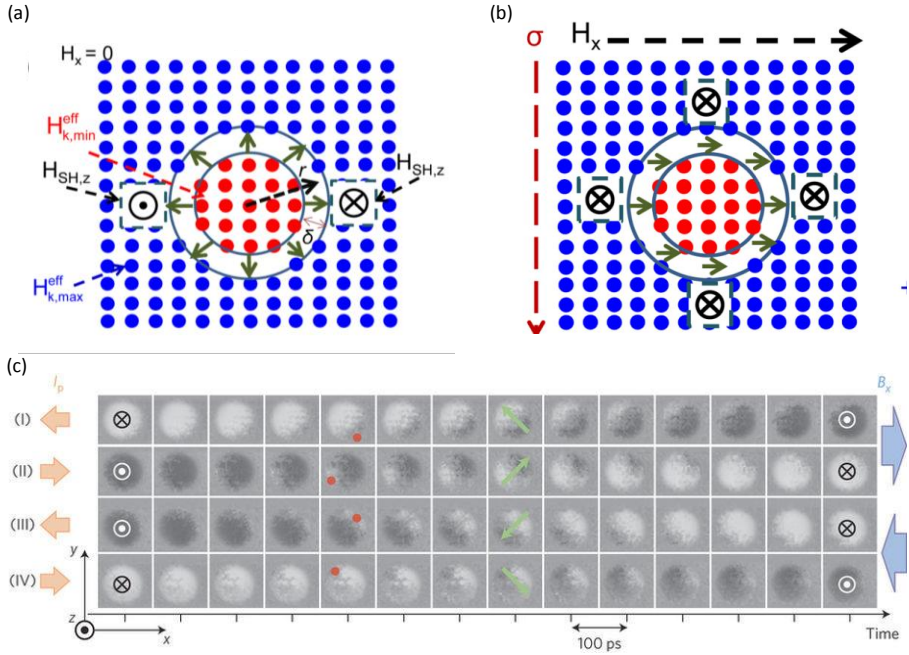

Fig. 8. (a) Schematic of a reversed domain with illustration of a chiral Néel domain wall stabilized by DMI for the case of zero assistive magnetic field. The direction of effective field due to the damping like SOT on the domain wall center is shown ($H_{SH,z}$). Reprinted figure with permission from O. J. Lee *et al.*, Phys. Rev. B **89**, 024418 (2014). Copyright (2014) by the American Physical Society. (b) Illustration of a reversed domain with a strong enough assistive field to overcome the DMI in the film. The SOT effective fields on the opposite edges of the domain are in the opposite directions for (a) leading to domain displacement and in the same direction for (b) leading to domain expansion. Reprinted figure with permission from O. J. Lee *et al.*, Phys. Rev. B **89**, 024418 (2014). Copyright (2014) by the American Physical Society. (c) Imaging of SOT switching in a Pt/Co/AlO$_X$ dot (500 nm diameter) for a 2 ns current pulse taken at 100 ps time step for the different directions of the current and assist field. The red dots indicate the domain nucleation edge and the green arrows indicate the direction of domain wall propagation during switching for each case. Reprinted by permission from Springer Customer Service Centre GmbH: Springer Nature, Nature Nanotechnology **12**, 980 (2017), M. Baumgartner *et al.*, Copyright 2017.
28

## C. Oscillatory switching behavior

In the previous sections, we discussed that $\boldsymbol{\tau}_{FL}$ plays a critical role in SOT switching dynamics in addition to $\boldsymbol{\tau}_{DL}$. Recently, it was found that a strong $\boldsymbol{\tau}_{FL}$ can result in oscillatory behavior of incoherent SOT switching.[145,146] Figure 9(a) shows the switching probability in a Ta/Co$_{40}$Fe$_{40}$B$_{20}$/MgO dot with a 1 μm diameter under short pulse current injection[146]. In contrast to Fig. 7(a), it is observed that in Fig. 9(a), as the pulse duration increases, the switching probability oscillates between 0% (blue regions) and 100% (white regions) for a given current density. A similar reduction in the switching probability after 100% switching was observed using time resolved MOKE experiments[145,147] in Ta/Co$_{40}$Fe$_{40}$B$_{20}$/MgO as shown in Fig. 9(b). By performing micromagnetic simulations, the switching back phenomenon was identified to arise from domain wall reflections at the sample edges. This domain wall reflection is illustrated in Fig. 9(c) and 9(d), which represent the magnetization configuration obtained for a pulse width of 1.7 ns and 1.8 ns, respectively, obtained from micromagnetic simulations. It is found that for a shorter pulse width, the magnetization switches from the initial red state to the blue state. For the case of larger pulse width domain walls are reflected and the magnetization switches back to initial red state. It was found that a strong $\boldsymbol{\tau}_{FL}$ plays a critical role in altering the domain wall dynamics to stabilize the reflected domain walls. Subsequently, the oscillatory switching behavior in Fig. 9(a) was also used to demonstrate a unipolar SOT switching scheme[146] which can be useful for applications to increase the scalability by replacing the driving transistors with diodes.

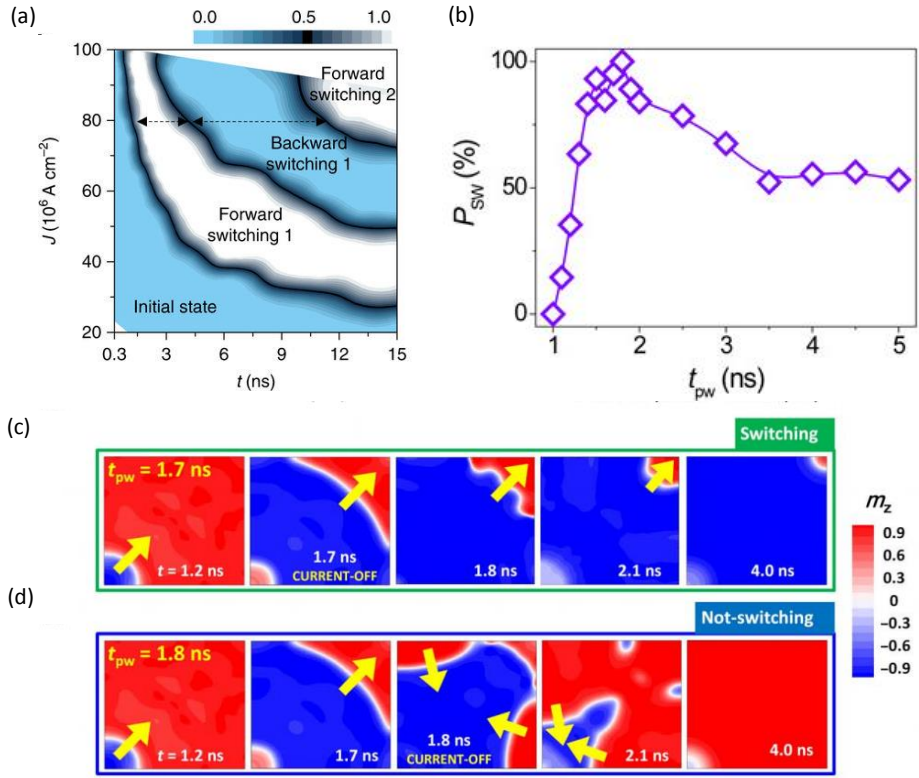

Fig. 9. (a) SOT induced magnetization switching probability under short pulse current injections in a Ta/Co$_{40}$Fe$_{40}$B$_{20}$/MgO dot for different applied pulse duration and current densities. Oscillatory behavior of the switching probability is observed. Reused from J. M. Lee *et al.*, Commun. Phys. **1**, 2 (2018), which is licensed under a Creative Commons Attribution 4.0 International License. (b) SOT switching probability measured from ps-MOKE for different pulse duration. A reduction in the switching probability for longer pulse duration indicates switching back phenomenon. Reused from J. Yoon *et al.*, Sci. Adv. **3**, e1603099 (2017), which is licensed under Creative Commons Attribution-NonCommercial license. (c,d) Micromagnetic simulation results for SOT switching for the pulse width of 1.7 ns (c) and 1.8 (d), which correspond to the cases of switching and switching back, respectively. Reused from J. Yoon *et al.*, Sci. Adv. **3**, e1603099 (2017), which is licensed under Creative Commons Attribution-NonCommercial license.

## IV. External field free SOT switching

As mentioned in Sec. I C, the deterministic magnetization switching in PMA structures using SOTs with in-plane spin polarization requires an external assist field to break the symmetry in the system. However, the incorporation of this assist field is not practical in real applications due to scalability reasons. The importance of this practical requirement to eliminate the external magnetic field was highlighted even in one of early works[14] of SOT switching, wherein the authors used the stray fields from the additional FMs deposited on the electrical contacts on the either side of the SOT device to achieve deterministic switching. While this proposed scheme is simple, depositing huge magnets on the either side of the device leads to scalability and uniformity issues. Various different approaches, such as structural and stack engineering, exchange biasing of a FM, the use of ferroelectric substrates and the use of geometrical domain wall pinning, were attempted to realize the deterministic external field free switching feature within the SOT device itself. In this section, we review the different experimental studies that have demonstrated external field free SOT switching in PMA structures and discuss on their feasibility from the perspective of practical implementations.

### A. Wedged structural engineering

The initial approaches to incorporate the deterministic SOT switching utilized wedging of the different stack components of the SOT device to break the symmetry in the system. Figure 10(a) illustrates the growth and patterning of a wedged oxide structure in a Ta (5 nm)/CoFeB (1 nm)/TaOx stack[155]. Due to the wedge, the level of oxidation and thus the strength of PMA varies along the lateral direction (y-axis in Fig. 10(a)) in the sample. Subsequently, the authors patterned a section of wedge into Hall bar devices as shown in Fig. 10(b) for electrical measurements. The authors find that in the devices with a lateral oxide gradient, apart from the conventional current induced SOT effective fields ($\mathbf{H}_y^{DL}$ and $\mathbf{H}_y^{FL}$), there exists an additional perpendicular field ($\mathbf{H}_z^{FL}$) which breaks the symmetry in system for the two magnetic states for a given current direction as shown in Fig. 10(c). Furthermore, the strength and the sign of this $\mathbf{H}_z^{FL}$ also depends on level of oxidation of the FM (opposite signs for over- and under-oxidation). Accordingly, for a given current direction, only one of the two magnetic states (up or down) is stable due to the symmetry breaking induced by $\mathbf{H}_z^{FL}$. Figure 10(d) illustrates a zero external magnetic field SOT switching in the lateral oxide structures for the case of under oxidation.

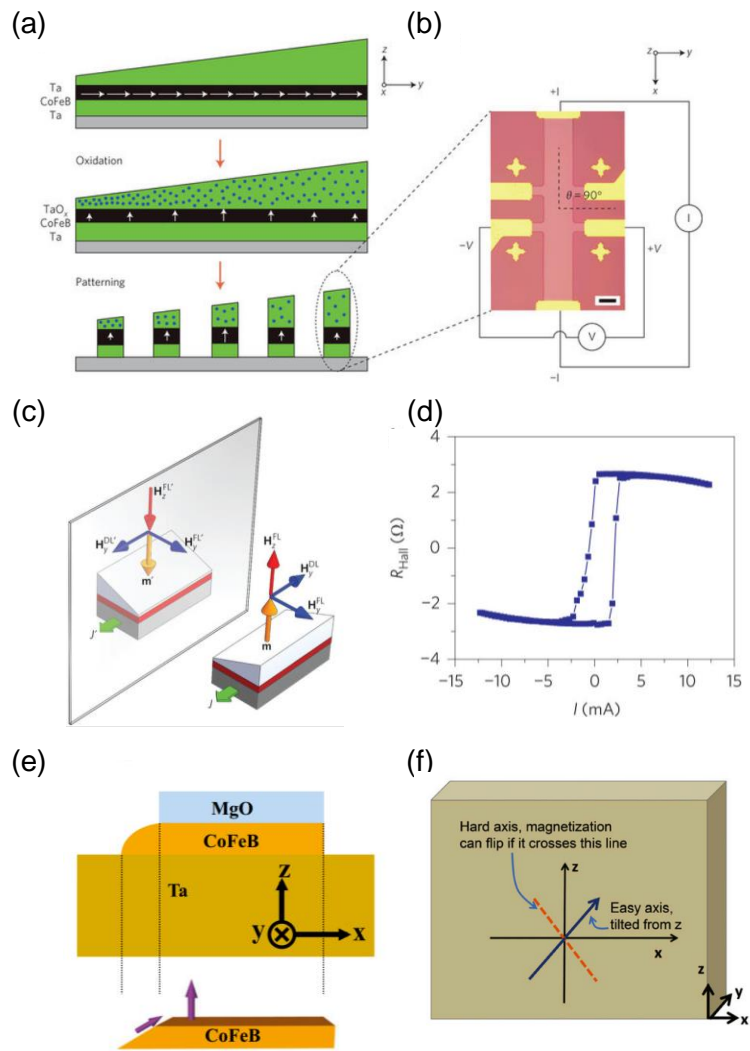

Fig. 10. (a) Growth and patterning steps to obtain a wedge shaped oxide on top of the FM. The wedge shaped structure results in a gradient of oxidation and thus PMA, along the wedge direction. Reprinted by permission from Springer Customer Service Centre GmbH: Springer Nature, Nature Nanotechnology **9**, 548 (2014), G. Yu *et al.*, Copyright 2014. (b) Device structure for electrical measurements patterned from the wedge structure in (a) and measurement configurations. Reprinted by permission from Springer Customer Service Centre GmbH: Springer Nature, Nature Nanotechnology **9**, 548 (2014), G. Yu *et al.*, Copyright 2014. (c) Illustration of mirror symmetry of different current induced torques in the wedged oxide structures. $\mathbf{H}_y^{DL}$ and $\mathbf{H}_y^{FL}$ are the effective fields corresponding to the two conventional components of current induced SOT $\boldsymbol{\tau}_{DL}$ and $\boldsymbol{\tau}_{FL}$, respectively, and $\mathbf{H}_y^{DL'}$ and $\mathbf{H}_y^{FL'}$ are their corresponding mirror reflections with respect to x-z plane. $\mathbf{H}_z^{FL}$ is the perpendicular field due to the oxide wedge structure. Reprinted by permission from Springer Customer Service Centre GmbH: Springer Nature, Nature Nanotechnology **9**, 548 (2014), G. Yu *et al.*, Copyright 2014. (d) Current induced switching using the wedged oxide structure without an external assistive field. Reprinted by permission from Springer Customer Service Centre GmbH: Springer Nature, Nature Nanotechnology **9**, 548 (2014), G. Yu *et al.*, Copyright 2014. (e) Schematic illustration of a wedge shaped CoFeB nanomagnet structure created by patterning techniques, resulting in a titled magnetic anisotropy of CoFeB. L. You *et al.*, Proc. Natl. Acad. Sci. U.S.A. **112**, 10310 (2015). (f) Schematic showing the orientation of easy and hard axes of the magnetic with its magnetic anisotropy slightly tilted away from the perpendicular z-axis. L. You *et al.*, Proc. Natl. Acad. Sci. U.S.A. **112**, 10310 (2015).

It is noted that, in this study, the authors explained the role of $\mathbf{H}_z^{FL}$ predominantly in a macroscopic picture. However, as we discussed in section III, the SOT switching in such large structures proceeds via. domain nucleation and propagation and the role of external assist fields is to overcome the DMI field in the structure[59]. Consequently, in a later work[156], the authors studied the microscopic picture of SOT switching in structures with lateral oxidation/anisotropy gradient due to the coexistence of $\mathbf{H}_z^{FL}$ and $\mathbf{H}_y^{DL}$, and found that, for the zero external assist field case the switching is indeed dominated by $\mathbf{H}_z^{FL}$ and for large external assist field case the switching is dominated by $\mathbf{H}_y^{DL}$. Furthermore, it is noted that the top oxide in these above works is TaOx that does not offer a high TMR ratio. To be incorporated into the SOT-MRAM scheme (Fig. 1(c)), the preferred top oxide is MgO. As a result, in a subsequent study[157], the authors demonstrated the external field free SOT switching in a Ta (5 nm)/CoFeB/MgO (5 nm)/TaOx stack by wedging the CoFeB layer (in contrast to TaOx in their previous works) and showed that the results still remain qualitatively similar. The above technique of wedging the FM layer to generate $\mathbf{H}_z^{FL}$ was extended to other material systems such as Hf/CoFeB/MgO[158], Hf/CoFeB/TaOx[158] and Pt/Co/MgO[159] in later works.

While these above works achieve the deterministic switching using the $\mathbf{H}_z^{FL}$ from lateral anisotropy gradient, Figure 10(e) illustrates an alternative wedging technique to achieve external field SOT switching.[160] The anisotropy in the region covered by MgO is PMA, while the anisotropy in the wedge region is not PMA due to the absence of MgO. Due to the requirements of minimization of magnetostatic energy, the magnetization in the wedge region follows the edge of the wedge as shown in Fig. 10(e), which results in a tilt in the easy axis of the nano-magnet as illustrated in Fig. 10(f). Subsequently, the authors showed that this tilting of anisotropy is sufficient to achieve zero external field deterministic magnetic switching and supported their conclusions using micromagnetic simulations. It is noted that in contrast to earlier works, there is no generation of $\mathbf{H}_z^{FL}$ in this work and the switching was attributed to be driven by $\boldsymbol{\tau}_{DL}$ alone. Following this work, it was found that such a tilt in anisotropy could also be created based on the relative position of the substrate and sputtering target.[161]

As a concluding remark, all these above techniques of engineering the magnetic anisotropy by creating wedged structures and growth techniques are interesting from the perspective of revealing rich physics of the symmetries in a magnetic system. However, the technique of wedging may be an obstacle for real device applications where wafer level homogeneity of magnetic and electrical properties is desired for mass production.

## B. Exchange coupling based techniques

The second class of techniques to eliminate the requirement of external assist fields in SOT switching utilized the exchange bias or interlayer coupling by the use of an AFM and/or additional FM layers to incorporate the assist field within the SOT device itself. As shown in Fig. 11, a variety of schemes were initially proposed to achieve deterministic field free SOT switching in PMA materials as well as to reveal the rich physics of antiferromagnetic SOT. In Figs. 11(a) and 11(b), the exchange bias fields arising from the AFMs, such as PtMn and IrMn, respectively adjacent to the FM, play the role of an assist field for SOT switching. Further, these AFMs themselves act as a source of SOT in both these cases. For the case of PtMn/[Co/Ni] structure[92] in Fig. 11(a), it was also found that by engineering the exchange bias in the system, the stack can show a memristor-like behavior and thus find applications in neuromorphic computing. The authors also showed in a later study[162] that proper engineering of the stack structure, such as changing the seed layer below PtMn from Pt to Ru, and inserting a thin Pt layer between Co/Ni and PtMn, can result in an enhanced PMA, enhanced exchange bias and reduced switching current density. In an another study[163], a composite FM of CoFeB/Gd/CoFeB with a reduced $M_S$ was used along with PtMn to demonstrate field free SOT switching. For the case of the structure[164] shown in Fig. 11(b), it was found that the bottom IPA CoFeB enhances the exchange coupling to achieve the complete field free switching. Furthermore, by performing various control experiments, the authors found that although IrMn serves as a SOT source the exact microscopic origins of SOT in IrMn in their system is not very clear and needs further exploration.

In the schemes shown in Figs. 11(c) and 11(d), the source of SOT arises from the bottom Pt layer. For the case in Fig. 11(c), the top IrMn layer provides the exchange bias field that was used to achieve deterministic external field less switching[165]. However, it was found that for this stack, the switching is not complete and was attributed to arise from the polycrystalline grain nature of IrMn. This suggests that the local crystalline structure of the AFM is important for achieving complete SOT switching. A similar structure as Fig. 11(c) with the Pd underlayer instead of Pt was studied, wherein IrMn acted as a source of exchange bias as well as SOT source.[166] For the case in Fig. 11(d), the magnetic field for external field-less SOT switching arises from the interlayer exchange coupling (IEC) via. the Ru layer[167]. By performing a Ru thickness dependence, the authors find that for smaller Ru thicknesses, the IEC from the top CoFe is very strong and overwhelms the PMA of the bottom free CoFe layer. Hence, a relatively large Ru thickness of 2 nm (IEC is antiferromagnetic) and 2.5 nm (IEC is ferromagnetic) were used to demonstrate external field-free SOT switching while retaining PMA. In a recent work[168], it was proposed that, by replacing the interlayer Ru with a material such as Ta with a weak IEC, the thickness of the interlayer can be reduced.

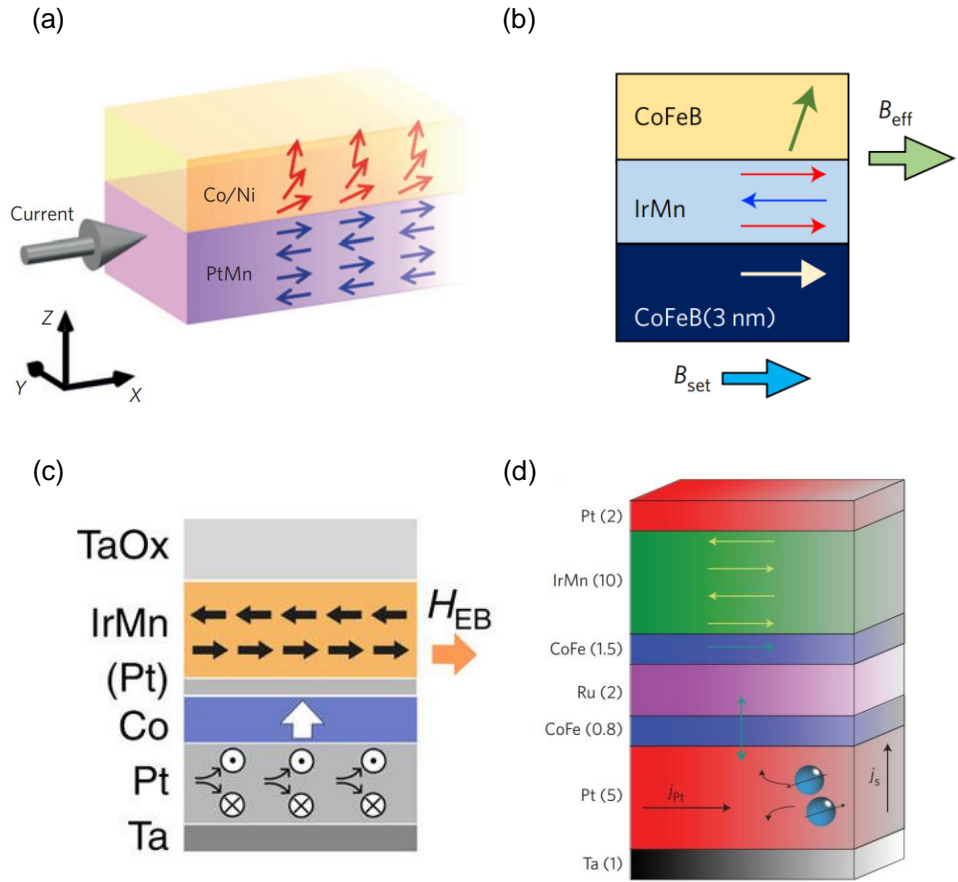

Fig. 11. Schematic stack structures of different external field free SOT switching schemes using exchange bias or interlayer exchange coupling. (a) PtMn serves as the exchange bias as well as SOT source. Reprinted by permission from Springer Customer Service Centre GmbH: Springer Nature, Nature Materials **15**, 535 (2016), S. Fukami *et al.*, Copyright 2016. (b) IrMn serves as an antiferromagnetic SOT source as well as exchange bias. Bottom CoFeB enhances the exchange bias. Reprinted by permission from Springer Customer Service Centre GmbH: Springer Nature, Nature Nanotechnology **11**, 878 (2016), Y.-W. Oh *et al.*, Copyright 2016. (c) IrMn provides an exchange bias field, the Pt dusting layer on top of Co enhances the PMA and the bottom Pt layer serves as a SOT source. Reused from A. van den Brink *et al.*, Nat. Commun. **7**, 10854 (2016), which is licensed under a Creative Commons Attribution 4.0 International License. (d) Interlayer

exchange via. Ru acts as an effective magnetic field for the bottom CoFe layer. The bottom Pt serves as a spin Hall source. Reprinted by permission from Springer Customer Service Centre GmbH: Springer Nature, Nature Nanotechnology **11**, 758 (2016), Y.-C. Lau *et al.*, Copyright 2016.

In terms of integration into three terminal SOT memories (Fig. 1(c)), it is not straightforward to integrate the structures shown in Figs. 11(c) and 11(d) as the PMA free layer is buried deep inside in the stack and cannot be interfaced directly with a tunnel barrier. However, they can be used for applications such as SOT oscillators. On the other hand, it is easier to integrate the structure in Figs. 11(a) and 11(b) as an MgO barrier can be interfaced on top of the free PMA layer. Apart from integration aspects, it is also necessary to consider the Joule heating as it was shown recently[169] that heating effects could decrease the exchange bias and degrade the external field free switching in AFM based SOT switching schemes. Moreover, repeated switching events can lead to degradation of the exchange bias between the AFM and FM layer and thus, affecting the long term stability of the memory device. Other than exchange coupling, it was shown[170] that a dipole coupling of the stray magnetic field from an in-plane FM could be also used to achieve an external field free SOT switching. It was proposed that, unlike the exchange biased schemes in Figs. 11(a) and 11(b) which require the AFM to be next to the free layer, the dipole coupling scheme can be used in stack with a conventional p-MTJ underlayer, such as Ta, below the free FM layer. Furthermore, the dipole coupling scheme also effectively eliminates the tradeoff between PMA and exchange bias, with respect to annealing conditions. A comparison of different magnetic parameters for some of the above external field free switching schemes can be found in the work[170].

## C. Electric field controlled SOT switching

So far, in the methods discussed to achieve the external field free switching, it is not possible to alter the direction of the effective magnetic field (wedged structured techniques) or there is a requirement of an external magnetic field cooling technique to change the direction of the effective magnetic field (AFM or exchange coupling technique). In a recent work[171], a field free SOT switching scheme using a ferroelectric/ferromagnetic hybrid structure was demonstrated where the direction of the effective magnetic field for SOT switching can be programmed using electric means. As shown in Fig. 12(a), Pt/Co/Ni/Co/Pt layers were deposited on a PMN-PT substrate and patterned for Hall bar measurements. A voltage $V_{PMN-PT}$ is applied to set the initial polarization of the ferroelectric substrate. As a result, a spin density gradient is generated when the current is injected into the channel of the SOT device resulting in a field free switching. As shown in Figs. 12(b) and 12(c), the direction of the switching loops can be controlled by changing the sign of the applied voltage to the substrate.

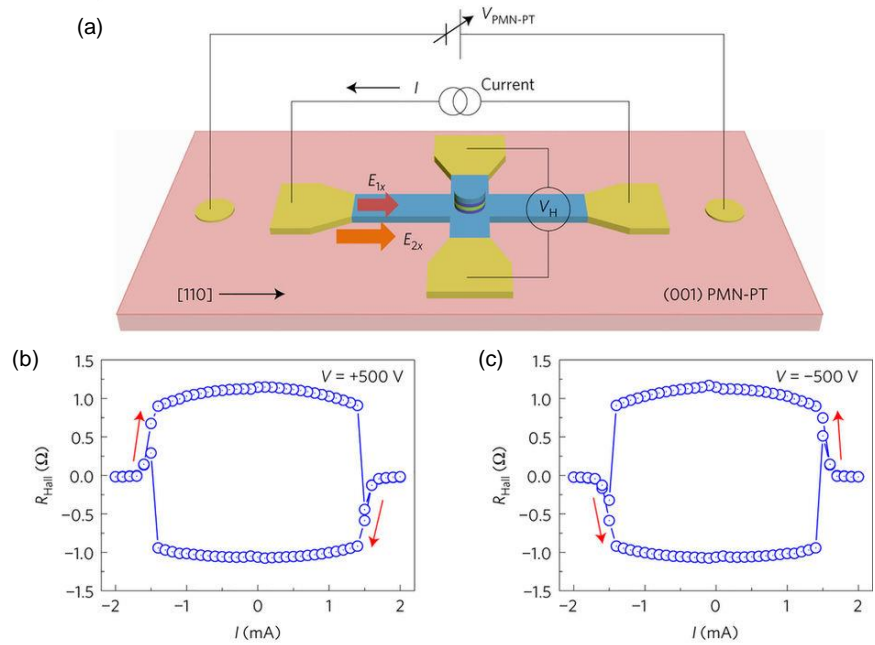

Fig. 12. (a) Schematic illustration of the measurement for the electric field control of deterministic SOT switching using a PMN-PT substrate. $V_{PMN-PT}$ is the applied voltage on the substrate to set its ferroelectric polarization and is removed during current induced switching measurements. Current induced switching loops obtained after applying a voltage of +500 V (b) and -500 V (c). Reprinted by permission from Springer Customer Service Centre GmbH: Springer Nature, Nature Materials **11**, 712 (2017), K. Cai *et al.*, Copyright 2017.

D. Out-of-plane spin polarization

The conventionally studied SOC effects generate in-plane polarized spins at the NM/FM interface when an in-plane charge current is injected. Very recently, research studies have attempted to generate an OOP spin polarization at the NM/FM interface using an in-plane charge current. In contrast to the case of the in-plane spins, the OOP spins can switch the magnetization of a FM without any assist field by exerting a SOT. For generation of this OOP spins, different approaches were followed, such as, the use of crystal symmetry[109], use of a FM/NM bilayer[172-175] as the spin current source, or even use a dual heavy metal bilayer[65] as the spin current source. Figure 13 shows the stack structures in which external field free switching was demonstrated using OOP spin generation. In Fig. 13(a), the interface between a FM (NiFe or CoFeB) and Ti is used to generate the OOP spin polarization[175]. By performing the hysteresis loop measurements and numerical simulations, the authors identified there is indeed an OOP spin polarization in the stack. In Fig. 13(b), the interface between two heavy metals with opposite spin Hall angle is used to demonstrate zero external field SOT switching[65]. The authors find that under appropriate thickness of W and Pt, the competing spin currents generated from the two layers generate an OOP effective field, which can subsequently switch the CoFeB layer with PMA without any external assistive magnetic field. The authors also note that the generation of this OOP field is beyond the current understanding of SOT and needs further exploration. In the stacks shown in Fig. 13, the FM with PMA can be interfaced with an MgO barrier in a straightforward manner and thus pose lesser engineering challenges towards incorporation into three terminal magnetic memories.

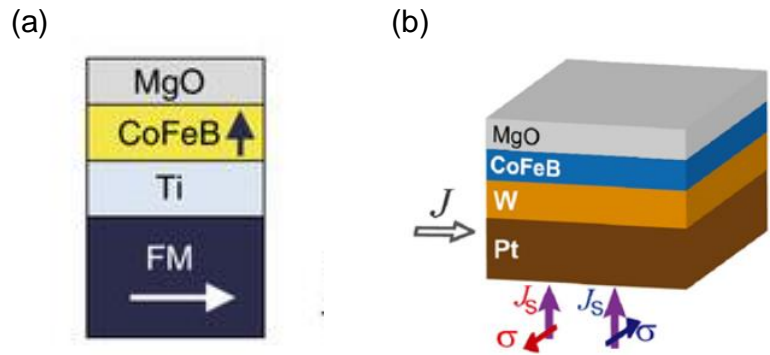

Fig. 13. Stack structure for generating OOP spins using (a) the FM/Ti interface (Reprinted by permission from Springer Customer Service Centre GmbH: Springer Nature, Nature Materials **17**, 509 (2018), S.-h. C. Baek *et al.*, Copyright 2018.), and (b) dual heavy metals with opposite spin Hall angle (Reprinted figure with permission from Q. Ma *et al.*, Phys. Rev. Lett. **120**, 117703 (2018). Copyright (2018) by the American Physical Society.).

**E. Geometrical domain-wall pinning**

Very recently, a new concept of external field free spin-orbit torque switching was demonstrated using the SOT driven domain wall motion in combination with geometrical domain wall pinning.[176] It is well known that the SOT with in-plane spins can drive domain walls in a perpendicular FM without an external magnetic field, in contrast to the case of switching. By designing anti-notched structures on the either end of a magnetic wire (see Fig. 14(a)), it was found that the SOT can move a domain wall back and forth in the wire by injecting currents of opposite polarities as illustrated in Fig. 14(b). The anti-notched structures serve as geometrical pinning sites to contain a single domain wall within the strip to achieve successive switching capabilities. The magnetization state in the center on the wire serves as the memory state and can be readout using a MTJ as illustrated in Fig. 14(a).

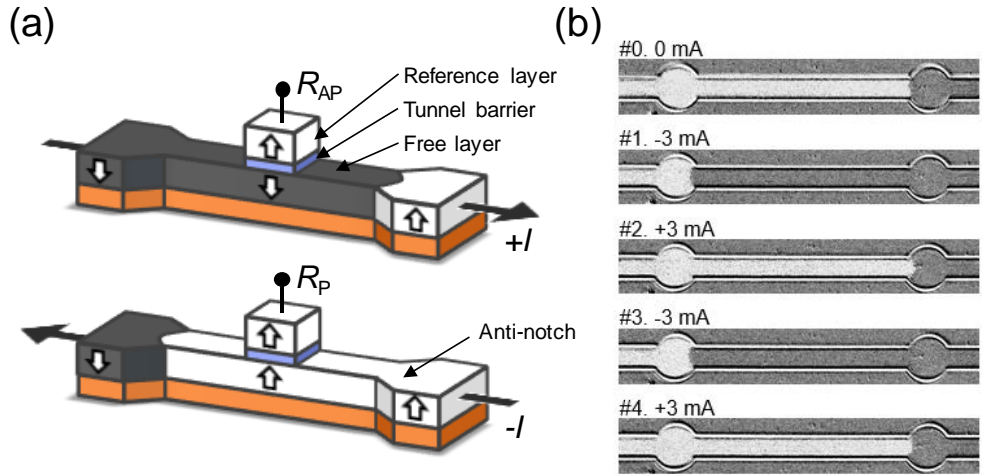

Fig. 14. (a) Schematic diagram of a field free three terminal magnetic memory composed of anti-notches and a MTJ. Adapted with permission from J. M. Lee *et al.*, Nano Lett., preprint doi:10.1021/acs.nanolett.8b00773 (2018). Copyright (2018) American Chemical Society. (b) Differentiated MOKE images of the magnetic microwire under a sequence of pulsed currents of opposite polarities (indicated on top left). The current is not applied for the initial measurement (I = 0 mA). The light and dark contrasts represent the two opposite magnetization states. Reprinted with permission from J. M. Lee *et al.*, Nano Lett., preprint doi:10.1021/acs.nanolett.8b00773 (2018). Copyright (2018) American Chemical Society.

## V.    Conclusions and outlooks

Spin-orbit torques have provided alternative and power efficient means to manipulate the magnetization compared to conventionally used spin transfer torques. As the read and write paths are decoupled, the SOT-MRAM offers flexibility in design margins and minimizes the chances of the tunnel breakdown issue unlike STT-MRAMs. On the other hand, in the SOT scheme, a high write current in the NM raises the chances of electromigration. As discussed in Sec. II, research is being carried out to improve the SOT efficiency and thus reduce the write current by techniques such as alloying and interface engineering. Furthermore, exotic materials, such as TIs are being explored to further reduce the SOT switching current. However, engineering issues of the exotic materials such as annealing stability and material compatibility with the existing Si platform need to be further researched. With respect to the FM layer in SOT devices, which is the data storing free layer in SOT-MRAMs, challenges remain to choose the optimum anisotropy and thickness of the free FM to meet the design tradeoff between a high thermal stability and low switching current densities. Novel materials such as multilayers, SAFs and ferrimagnets can help in this regard. A summary for various material choices for the different layers of the SOT device is presented in Fig. 15. The strength of SOT can be tuned by a variety of material choices according to the design requirements.

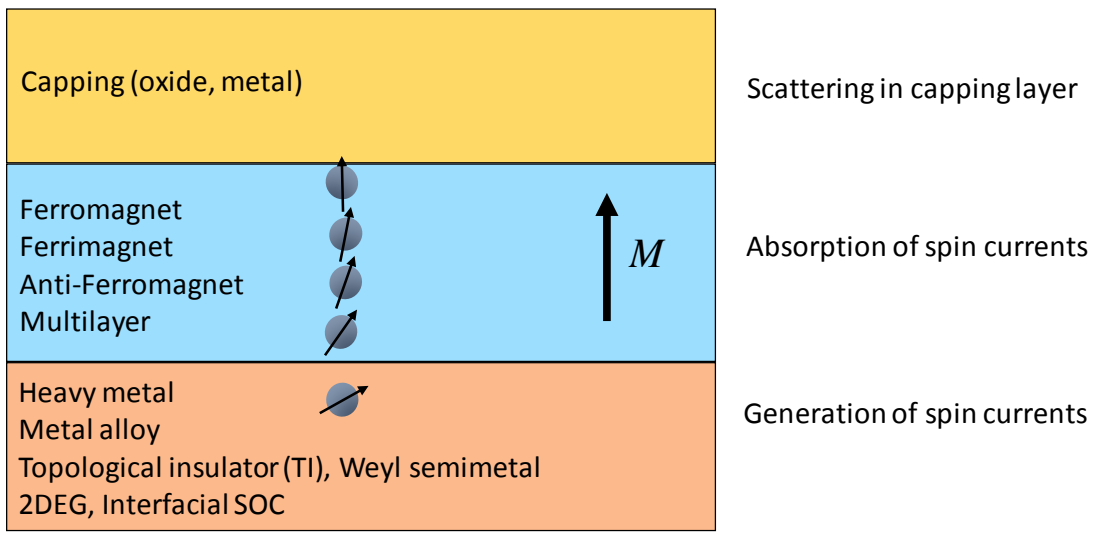

Fig. 15. Various material choices for different layers of a SOT device.

In Sec. III, we discussed about the spatial and time-resolved measurements, which revealed that SOT switching predominantly follows domain nucleation and expansion process. In addition to the damping-like torque, it was found that the field-like torque also plays a dominant role in SOT switching dynamics and can even cause a backward switching. Nevertheless, the incubation delay at the onset of the SOT switching process was measured to be negligible. In terms of switching studies, single shot electrical measurements of the SOT switching process are yet to be explored. The requirement of an external magnetic field to achieve deterministic switching harms the scalability of the SOT devices. As we discussed in Sec. IV, many techniques such as the wedged structuring, use of exchange and dipolar coupling, and use of ferroelectric substrates have been proposed to incorporate the assistive field within the SOT device itself. Recent studies have also attempted to generate an OOP spin polarization from in-plane currents or utilize geometrical domain wall pinning to achieve deterministic switching of a PMA magnet without an assist field.

This review serves as a brief summary focusing on the latest advancements in SOT researches. Apart from the SOT studies aiming to understand the underlying SOC physics, research is still required for tackling the engineering issues to implement SOT devices into practical applications. For example, more studies focusing on the reliability of the SOT devices and SOT switching error rates by performing endurance tests would be some important avenues from the perspective of the industry applications.

This research is supported by the National Research Foundation, Prime Minister's Office, Singapore, under its Competitive Research Programme (CRP Grant No. NRF CRP12-2013-01).

# References

1. W. Thomson, Proc. Royal Soc. Lond. **8,** 546 (1857).
2. M. N. Baibich, J. M. Broto, A. Fert, F. N. Van Dau, F. Petroff, P. Etienne, G. Creuzet, A. Friederich, and J. Chazelas, Phys. Rev. Lett. **61,** 2472 (1988).
3. J. Barnaś, A. Fuss, R. E. Camley, P. Grünberg, and W. Zinn, Phys. Rev. B **42,** 8110 (1990).
4. M. Julliere, Phys. Lett. A **54,** 225 (1975).
5. J. S. Moodera, L. R. Kinder, T. M. Wong, and R. Meservey, Phys. Rev. Lett. **74,** 3273 (1995).
6. T. Miyazaki and N. Tezuka, J. Magn. Magn. Mater. **139,** L231 (1995).
7. S. Parkin, J. Xin, C. Kaiser, A. Panchula, K. Roche, and M. Samant, Proc. IEEE **91,** 661 (2003).
8. L. Berger, Phys. Rev. B **54,** 9353 (1996).
9. J. C. Slonczewski, J. Magn. Magn. Mater. **159,** L1 (1996).
10. T. Min, Q. Chen, R. Beach, G. Jan, C. Horng, W. Kula, T. Torng, R. Tong, T. Zhong, D. Tang, P. Wang, M. m. Chen, J. Z. Sun, J. K. Debrosse, D. C. Worledge, T. M. Maffitt, and W. J. Gallagher, IEEE Trans. Magn. **46,** 2322 (2010).
11. W. S. Zhao, Y. Zhang, T. Devolder, J. O. Klein, D. Ravelosona, C. Chappert, and P. Mazoyer, Microelectron. Reliab. **52,** 1848 (2012).
12. A. Chernyshov, M. Overby, X. Liu, J. K. Furdyna, Y. Lyanda-Geller, and L. P. Rokhinson, Nat. Phys. **5,** 656 (2009).
13. I. Mihai Miron, G. Gaudin, S. Auffret, B. Rodmacq, A. Schuhl, S. Pizzini, J. Vogel, and P. Gambardella, Nat. Mater. **9,** 230 (2010).
14. I. M. Miron, K. Garello, G. Gaudin, P.-J. Zermatten, M. V. Costache, S. Auffret, S. Bandiera, B. Rodmacq, A. Schuhl, and P. Gambardella, Nature **476,** 189 (2011).
15. L. Liu, C. F. Pai, Y. Li, H. W. Tseng, D. C. Ralph, and R. A. Buhrman, Science **336,** 555 (2012).
16. L. Liu, O. J. Lee, T. J. Gudmundsen, D. C. Ralph, and R. A. Buhrman, Phys. Rev. Lett. **109,** 096602 (2012).
17. L. Q. Liu, T. Moriyama, D. C. Ralph, and R. A. Buhrman, Phys. Rev. Lett. **106,** 036601 (2011).
18. S. V. Aradhya, G. E. Rowlands, J. Oh, D. C. Ralph, and R. A. Buhrman, Nano Lett. **16,** 5987 (2016).
19. K. Garello, C. O. Avci, I. M. Miron, M. Baumgartner, A. Ghosh, S. Auffret, O. Boulle, G. Gaudin, and P. Gambardella, Appl. Phys. Lett. **105,** 212402 (2014).
20. S.-H. Yang, K.-S. Ryu, and S. Parkin, Nat. Nanotechnol. **10,** 221 (2015).
21. M. I. Dyakonov and V. I. Perel, Phys. Lett. A **35,** 459 (1971).
22. J. E. Hirsch, Phys. Rev. Lett. **83,** 1834 (1999).
23. Y. K. Kato, R. C. Myers, A. C. Gossard, and D. D. Awschalom, Science **306,** 1910 (2004).
24. S. O. Valenzuela and M. Tinkham, Nature **442,** 176 (2006).
25. E. Saitoh, M. Ueda, H. Miyajima, and G. Tatara, Appl. Phys. Lett. **88,** 182509 (2006).
26. A. Hoffmann, IEEE Trans. Magn. **49,** 5172 (2013).
27. J. Sinova, S. O. Valenzuela, J. Wunderlich, C. H. Back, and T. Jungwirth, Rev. Mod. Phys. **87,** 1213 (2015).
28. V. M. Edelstein, Solid State Commun. **73,** 233 (1990).
29. G. Dresselhaus, Phys. Rev. **100,** 580 (1955).
30. Y. A. Bychkov and E. I. Rashba, JETP Lett. **39,** 78 (1984).
31. I. M. Miron, T. Moore, H. Szambolics, L. D. Buda-Prejbeanu, S. Auffret, B. Rodmacq, S. Pizzini, J. Vogel, M. Bonfim, A. Schuhl, and G. Gaudin, Nat. Mater. **10,** 419 (2011).
32. A. Manchon and S. Zhang, Phys. Rev. B **79,** 094422 (2009).
33. J. C. R. Sánchez, L. Vila, G. Desfonds, S. Gambarelli, J. P. Attané, J. M. De Teresa, C. Magén, and A. Fert, Nat. Commun. **4,** 2944 (2013).
34. C. Ciccarelli, K. M. D. Hals, A. Irvine, V. Novak, Y. Tserkovnyak, H. Kurebayashi, A. Brataas, and A. Ferguson, Nat. Nanotechnol. **10,** 50 (2014).
35. A. Manchon, H. C. Koo, J. Nitta, S. M. Frolov, and R. A. Duine, Nat. Mater. **14,** 871 (2015).

36 K.-W. Kim, S.-M. Seo, J. Ryu, K.-J. Lee, and H.-W. Lee, Phys. Rev. B **85,** 180404 (2012).
37 A. Manchon, "Spin Hall effect versus Rashba torque: a Diffusive Approach", preprint arXiv:1204.4869 (2012).
38 P. M. Haney, H.-W. Lee, K.-J. Lee, A. Manchon, and M. D. Stiles, Phys. Rev. B **87,** 174411 (2013).
39 J. Kim, J. Sinha, M. Hayashi, M. Yamanouchi, S. Fukami, T. Suzuki, S. Mitani, and H. Ohno, Nat. Mater. **12,** 240 (2013).
40 K. Garello, I. M. Miron, C. O. Avci, F. Freimuth, Y. Mokrousov, S. Blugel, S. Auffret, O. Boulle, G. Gaudin, and P. Gambardella, Nat. Nanotechnol. **8,** 587 (2013).
41 X. Qiu, P. Deorani, K. Narayanapillai, K.-S. Lee, K.-J. Lee, H.-W. Lee, and H. Yang, Sci. Rep. **4,** 4491 (2014).
42 M. Hayashi, J. Kim, M. Yamanouchi, and H. Ohno, Phys. Rev. B **89,** 144425 (2014).
43 U. H. Pi, K. Won Kim, J. Y. Bae, S. C. Lee, Y. J. Cho, K. S. Kim, and S. Seo, Appl. Phys. Lett. **97,** 162507 (2010).
44 C. O. Avci, K. Garello, M. Gabureac, A. Ghosh, A. Fuhrer, S. F. Alvarado, and P. Gambardella, Phys. Rev. B **90,** 224427 (2014).
45 A. R. Mellnik, J. S. Lee, A. Richardella, J. L. Grab, P. J. Mintun, M. H. Fischer, A. Vaezi, A. Manchon, E. A. Kim, N. Samarth, and D. C. Ralph, Nature **511,** 449 (2014).
46 D. Fang, H. Kurebayashi, J. Wunderlich, K. Výborný, L. P. Zârbo, R. P. Campion, A. Casiraghi, B. L. Gallagher, T. Jungwirth, and A. J. Ferguson, Nat. Nanotechnol. **6,** 413 (2011).
47 X. Fan, J. Wu, Y. Chen, M. J. Jerry, H. Zhang, and J. Q. Xiao, Nat. Commun. **4,** 1799 (2013).
48 X. Fan, H. Celik, J. Wu, C. Ni, K.-J. Lee, V. O. Lorenz, and J. Q. Xiao, Nat. Commun. **5,** 3042 (2014).
49 M. Montazeri, P. Upadhyaya, M. C. Onbasli, G. Yu, K. L. Wong, M. Lang, Y. Fan, X. Li, P. Khalili Amiri, R. N. Schwartz, C. A. Ross, and K. L. Wang, Nat. Commun. **6,** 8958 (2015).
50 M. Cubukcu, O. Boulle, N. Mikuszeit, C. Hamelin, T. Brächer, N. Lamard, M. C. Cyrille, L. Buda-Prejbeanu, K. Garello, I. M. Miron, O. Klein, G. d. Loubens, V. V. Naletov, J. Langer, B. Ocker, P. Gambardella, and G. Gaudin, IEEE Trans. Magn. **54,** 1 (2018).
51 S. Mangin, D. Ravelosona, J. A. Katine, M. J. Carey, B. D. Terris, and E. E. Fullerton, Nat. Mater. **5,** 210 (2006).
52 K. S. Lee, S. W. Lee, B. C. Min, and K. J. Lee, Appl. Phys. Lett. **102,** 112410 (2013).
53 S. Yan and Y. B. Bazaliy, Phys. Rev. B **91,** 214424 (2015).
54 T. Taniguchi, S. Mitani, and M. Hayashi, Phys. Rev. B **92,** 024428 (2015).
55 R. Ramaswamy, X. Qiu, T. Dutta, S. D. Pollard, and H. Yang, Appl. Phys. Lett. **108,** 202406 (2016).
56 X. P. Qiu, K. Narayanapillai, Y. Wu, P. Deorani, D. H. Yang, W. S. Noh, J. H. Park, K. J. Lee, H. W. Lee, and H. Yang, Nat. Nanotechnol. **10,** 333 (2015).
57 J. Yu, X. Qiu, W. Legrand, and H. Yang, Appl. Phys. Lett. **109,** 042403 (2016).
58 L. M. Loong, P. Deorani, X. Qiu, and H. Yang, Appl. Phys. Lett. **107,** 022405 (2015).
59 O. J. Lee, L. Q. Liu, C. F. Pai, Y. Li, H. W. Tseng, P. G. Gowtham, J. P. Park, D. C. Ralph, and R. A. Buhrman, Phys. Rev. B **89,** 024418 (2014).
60 R. Mishra, J. Yu, X. Qiu, M. Motapothula, T. Venkatesan, and H. Yang, Phys. Rev. Lett. **118,** 167201 (2017).
61 C. F. Pai, L. Q. Liu, Y. Li, H. W. Tseng, D. C. Ralph, and R. A. Buhrman, Appl. Phys. Lett. **101,** 122404 (2012).
62 J. Torrejon, J. Kim, J. Sinha, S. Mitani, M. Hayashi, M. Yamanouchi, and H. Ohno, Nat. Commun. **5,** 4655 (2014).
63 M. Akyol, J. G. Alzate, G. Q. Yu, P. Upadhyaya, K. L. Wong, A. Ekicibil, P. K. Amiri, and K. L. Wang, Appl. Phys. Lett. **106,** 032406 (2015).
64 P. He, X. Qiu, V. L. Zhang, Y. Wu, M. H. Kuok, and H. Yang, Adv. Electron. Mater. **2,** 1600210 (2016).
65 Q. Ma, Y. Li, D. B. Gopman, Y. P. Kabanov, R. D. Shull, and C. L. Chien, Phys. Rev. Lett. **120,** 117703 (2018).
66 M.-H. Nguyen, C.-F. Pai, K. X. Nguyen, D. A. Muller, D. C. Ralph, and R. A. Buhrman, Appl. Phys. Lett. **106,** 222402 (2015).

67 C.-F. Pai, M.-H. Nguyen, C. Belvin, L. H. Vilela-Leão, D. C. Ralph, and R. A. Buhrman, Appl. Phys. Lett. **104,** 082407 (2014).
68 S. Woo, M. Mann, A. J. Tan, L. Caretta, and G. S. D. Beach, Appl. Phys. Lett. **105,** 212404 (2014).
69 J. Yu, X. Qiu, Y. Wu, J. Yoon, P. Deorani, J. M. Besbas, A. Manchon, and H. Yang, Sci. Rep. **6,** 32629 (2016).
70 X. Qiu, W. Legrand, P. He, Y. Wu, J. Yu, R. Ramaswamy, A. Manchon, and H. Yang, Phys. Rev. Lett. **117,** 217206 (2016).
71 J. Smit, Physica **24,** 39 (1958).
72 L. Berger, Phys. Rev. B **2,** 4559 (1970).
73 Y. Niimi, H. Suzuki, Y. Kawanishi, Y. Omori, T. Valet, A. Fert, and Y. Otani, Phys. Rev. B **89,** 054401 (2014).
74 Y. Niimi, Y. Kawanishi, D. H. Wei, C. Deranlot, H. X. Yang, M. Chshiev, T. Valet, A. Fert, and Y. Otani, Phys. Rev. Lett. **109,** 156602 (2012).
75 Y. Niimi, M. Morota, D. H. Wei, C. Deranlot, M. Basletic, A. Hamzic, A. Fert, and Y. Otani, Phys. Rev. Lett. **106,** 126601 (2011).
76 M. Yamanouchi, L. Chen, J. Kim, M. Hayashi, H. Sato, S. Fukami, S. Ikeda, F. Matsukura, and H. Ohno, Appl. Phys. Lett. **102,** 212408 (2013).
77 S. Takizawa, M. Kimata, Y. Omori, Y. Niimi, and Y. Otani, Appl. Phys. Express **9,** 063009 (2016).
78 R. Ramaswamy, Y. Wang, M. Elyasi, M. Motapothula, T. Venkatesan, X. Qiu, and H. Yang, Phys. Rev. Appl. **8,** 024034 (2017).
79 L. K. Zou, S. H. Wang, Y. Zhang, J. R. Sun, J. W. Cai, and S. S. Kang, Phys. Rev. B **93,** 014422 (2016).
80 J. Wu, L. Zou, T. Wang, Y. Chen, J. Cai, J. Hu, and J. Q. Xiao, IEEE Trans. Magn. **52,** 1 (2016).
81 Y. Wen, J. Wu, P. Li, Q. Zhang, Y. Zhao, A. Manchon, J. Q. Xiao, and X. Zhang, Phys. Rev. B **95,** 104403 (2017).
82 Y. Niimi and Y. Otani, Rep. Prog. Phys. **78,** 124501 (2015).
83 P. Laczkowski, J. C. Rojas-Sanchez, W. Savero-Torres, H. Jaffres, N. Reyren, C. Deranlot, L. Notin, C. Beigne, A. Marty, J. P. Attane, L. Vila, J. M. George, and A. Fert, Appl. Phys. Lett. **104,** 142403 (2014).
84 P. Laczkowski, Y. Fu, H. Yang, J. C. Rojas-Sánchez, P. Noel, V. T. Pham, G. Zahnd, C. Deranlot, S. Collin, C. Bouard, P. Warin, V. Maurel, M. Chshiev, A. Marty, J. P. Attané, A. Fert, H. Jaffrès, L. Vila, and J. M. George, Phys. Rev. B **96,** 140405 (2017).
85 M.-H. Nguyen, M. Zhao, D. C. Ralph, and R. A. Buhrman, Appl. Phys. Lett. **108,** 242407 (2016).
86 M.-H. Nguyen, S. Shi, G. E. Rowlands, S. V. Aradhya, C. L. Jermain, D. C. Ralph, and R. A. Buhrman, Appl. Phys. Lett. **112,** 062404 (2018).
87 D. Qu, S. Y. Huang, G. Y. Guo, and C. L. Chien, Phys. Rev. B **97,** 024402 (2018).
88 K.-U. Demasius, T. Phung, W. Zhang, B. P. Hughes, S.-H. Yang, A. Kellock, W. Han, A. Pushp, and S. S. P. Parkin, Nat. Commun. **7,** 10644 (2016).
89 H. An, Y. Kageyama, Y. Kanno, N. Enishi, and K. Ando, Nat. Commun. **7,** 13069 (2016).
90 J. B. S. Mendes, R. O. Cunha, O. Alves Santos, P. R. T. Ribeiro, F. L. A. Machado, R. L. Rodríguez-Suárez, A. Azevedo, and S. M. Rezende, Phys. Rev. B **89,** 140406 (2014).
91 W. Zhang, M. B. Jungfleisch, W. Jiang, J. E. Pearson, A. Hoffmann, F. Freimuth, and Y. Mokrousov, Phys. Rev. Lett. **113,** 196602 (2014).
92 S. Fukami, C. Zhang, S. DuttaGupta, A. Kurenkov, and H. Ohno, Nat. Mater. **15,** 535 (2016).
93 J. Sklenar, W. Zhang, M. B. Jungfleisch, W. Jiang, H. Saglam, J. E. Pearson, J. B. Ketterson, and A. Hoffmann, AIP Advances **6,** 055603 (2016).
94 W. Zhang, M. B. Jungfleisch, F. Freimuth, W. Jiang, J. Sklenar, J. E. Pearson, J. B. Ketterson, Y. Mokrousov, and A. Hoffmann, Phys. Rev. B **92,** 144405 (2015).
95 W. Zhang, W. Han, S.-H. Yang, Y. Sun, Y. Zhang, B. Yan, and S. S. P. Parkin, Sci. Adv. **2,** e1600759 (2016).
96 Y. Fan, P. Upadhyaya, X. Kou, M. Lang, S. Takei, Z. Wang, J. Tang, L. He, L.-T. Chang, M. Montazeri, G. Yu, W. Jiang, T. Nie, R. N. Schwartz, Y. Tserkovnyak, and K. L. Wang, Nat. Mater. **13,** 699 (2014).

97  Y. Wang, P. Deorani, K. Banerjee, N. Koirala, M. Brahlek, S. Oh, and H. Yang, Phys. Rev. Lett. **114,** 257202 (2015).
98  Y. Wang, D. Zhu, Y. Wu, Y. Yang, J. Yu, R. Ramaswamy, R. Mishra, S. Shi, M. Elyasi, K.-L. Teo, Y. Wu, and H. Yang, Nat. Commun. **8,** 1364 (2017).
99  K. Kondou, R. Yoshimi, A. Tsukazaki, Y. Fukuma, J. Matsuno, K. S. Takahashi, M. Kawasaki, Y. Tokura, and Y. Otani, Nat. Phys. **12,** 1027 (2016).
100  P. Deorani, J. Son, K. Banerjee, N. Koirala, M. Brahlek, S. Oh, and H. Yang, Phys. Rev. B **90,** 094403 (2014).
101  Y. Shiomi, K. Nomura, Y. Kajiwara, K. Eto, M. Novak, K. Segawa, Y. Ando, and E. Saitoh, Phys. Rev. Lett. **113,** 196601 (2014).
102  H. Wang, J. Kally, J. S. Lee, T. Liu, H. Chang, D. R. Hickey, K. A. Mkhoyan, M. Wu, A. Richardella, and N. Samarth, Phys. Rev. Lett. **117,** 076601 (2016).
103  L. Liu, A. Richardella, I. Garate, Y. Zhu, N. Samarth, and C.-T. Chen, Phys. Rev. B **91,** 235437 (2015).
104  J. S. Lee, A. Richardella, D. R. Hickey, K. A. Mkhoyan, and N. Samarth, Phys. Rev. B **92,** 155312 (2015).
105  M. Jamali, J. S. Lee, J. S. Jeong, F. Mahfouzi, Y. Lv, Z. Zhao, B. K. Nikolić, K. A. Mkhoyan, N. Samarth, and J.-P. Wang, Nano Lett. **15,** 7126 (2015).
106  Q. Shao, G. Yu, Y.-W. Lan, Y. Shi, M.-Y. Li, C. Zheng, X. Zhu, L.-J. Li, P. K. Amiri, and K. L. Wang, Nano Lett. **16,** 7514 (2016).
107  W. Zhang, J. Sklenar, B. Hsu, W. Jiang, M. B. Jungfleisch, J. Xiao, F. Y. Fradin, Y. Liu, J. E. Pearson, J. B. Ketterson, Z. Yang, and A. Hoffmann, APL Materials **4,** 032302 (2016).
108  J. Sklenar, W. Zhang, M. B. Jungfleisch, W. Jiang, H. Saglam, J. E. Pearson, J. B. Ketterson, and A. Hoffmann, J. Appl. Phys. **120,** 180901 (2016).
109  D. MacNeill, G. M. Stiehl, M. H. D. Guimaraes, R. A. Buhrman, J. Park, and D. C. Ralph, Nat. Phys. **13,** 300 (2016).
110  D. MacNeill, G. M. Stiehl, M. H. D. Guimarães, N. D. Reynolds, R. A. Buhrman, and D. C. Ralph, Phys. Rev. B **96,** 054450 (2017).
111  E. Lesne, Y. Fu, S. Oyarzun, J. C. Rojas-Sánchez, D. C. Vaz, H. Naganuma, G. Sicoli, J. P. Attané, M. Jamet, E. Jacquet, J. M. George, A. Barthélémy, H. Jaffrès, A. Fert, M. Bibes, and L. Vila, Nat. Mater. **15,** 1261 (2016).
112  J. Y. Chauleau, M. Boselli, S. Gariglio, R. Weil, G. d. Loubens, J. M. Triscone, and M. Viret, EPL **116,** 17006 (2016).
113  Y. Wang, R. Ramaswamy, M. Motapothula, K. Narayanapillai, D. Zhu, J. Yu, T. Venkatesan, and H. Yang, Nano Lett. **17,** 7659 (2017).
114  Q. Song, H. Zhang, T. Su, W. Yuan, Y. Chen, W. Xing, J. Shi, J. Sun, and W. Han, Sci. Adv. **3,** e1602312 (2017).
115  K. Narayanapillai, K. Gopinadhan, X. Qiu, A. Annadi, Ariando, T. Venkatesan, and H. Yang, Appl. Phys. Lett. **105,** 162405 (2014).
116  J. Han, A. Richardella, S. A. Siddiqui, J. Finley, N. Samarth, and L. Liu, Phys. Rev. Lett. **119,** 077702 (2017).
117  M. DC, M. Jamali, J.-Y. Chen, D. R. Hickey, D. Zhang, Z. Zhao, H. Li, P. Quarterman, Y. Lv, M. Li, K. A. Mkhoyan, and J.-P. Wang, "Room-temperature perpendicular magnetization switching through giant spin-orbit torque from sputtered BiXSe1-x topological insulator material", preprint arXiv:1703.03822 (2017).
118  M.-J. Jin, S. Y. Moon, J. Park, V. Modepalli, J. Jo, S.-I. Kim, H. C. Koo, B.-C. Min, H.-W. Lee, S.-H. Baek, and J.-W. Yoo, Nano Lett. **17,** 36 (2017).
119  R. Ohshima, Y. Ando, K. Matsuzaki, T. Susaki, M. Weiler, S. Klingler, H. Huebl, E. Shikoh, T. Shinjo, S. T. B. Goennenwein, and M. Shiraishi, Nat. Mater. **16,** 609 (2017).
120  S. Li, S. Goolaup, J. Kwon, F. Luo, W. Gan, and W. S. Lew, Sci. Rep. **7,** 972 (2017).
121  M. Jamali, K. Narayanapillai, X. Qiu, L. M. Loong, A. Manchon, and H. Yang, Phys. Rev. Lett. **111,** 246602 (2013).
122  K.-F. Huang, D.-S. Wang, H.-H. Lin, and C.-H. Lai, Appl. Phys. Lett. **107,** 232407 (2015).
123  C. Bi, H. Almasi, K. Price, T. Newhouse-Illige, M. Xu, S. R. Allen, X. Fan, and W. Wang, Phys. Rev. B **95,** 104434 (2017).

124  P. Wadley, B. Howells, J. Železný, C. Andrews, V. Hills, R. P. Campion, V. Novák, K. Olejník, F. Maccherozzi, S. S. Dhesi, S. Y. Martin, T. Wagner, J. Wunderlich, F. Freimuth, Y. Mokrousov, J. Kuneš, J. S. Chauhan, M. J. Grzybowski, A. W. Rushforth, K. W. Edmonds, B. L. Gallagher, and T. Jungwirth, Science **351,** 587 (2016).

125  Y. Yang, Y. Xu, X. Zhang, Y. Wang, S. Zhang, R.-W. Li, M. S. Mirshekarloo, K. Yao, and Y. Wu, Phys. Rev. B **93,** 094402 (2016).

126  J. Finley and L. Liu, Phys. Rev. Appl. **6,** 054001 (2016).

127  N. Roschewsky, C.-H. Lambert, and S. Salahuddin, Phys. Rev. B **96,** 064406 (2017).

128  K. Ueda, M. Mann, P. W. P. de Brouwer, D. Bono, and G. S. D. Beach, Phys. Rev. B **96,** 064410 (2017).

129  S.-G. Je, J.-C. Rojas-Sánchez, T. H. Pham, P. Vallobra, G. Malinowski, D. Lacour, T. Fache, M.-C. Cyrille, D.-Y. Kim, S.-B. Choe, M. Belmeguenai, M. Hehn, S. Mangin, G. Gaudin, and O. Boulle, Appl. Phys. Lett. **112,** 062401 (2018).

130  W. S. Ham, S. Kim, D.-H. Kim, K.-J. Kim, T. Okuno, H. Yoshikawa, A. Tsukamoto, T. Moriyama, and T. Ono, Appl. Phys. Lett. **110,** 242405 (2017).

131  J. Kim, D. Lee, K.-J. Lee, B.-K. Ju, H. C. Koo, B.-C. Min, and O. Lee, Sci. Rep. **8,** 6017 (2018).

132  Z. Zhao, M. Jamali, A. K. Smith, and J.-P. Wang, Appl. Phys. Lett. **106,** 132404 (2015).

133  D. Bang, J. Yu, X. Qiu, Y. Wang, H. Awano, A. Manchon, and H. Yang, Phys. Rev. B **93,** 174424 (2016).

134  K.-S. Lee, S.-W. Lee, B.-C. Min, and K.-J. Lee, Appl. Phys. Lett. **104,** 072413 (2014).

135  W. Legrand, R. Ramaswamy, R. Mishra, and H. Yang, Phys. Rev. Appl. **3,** 064012 (2015).

136  J. Park, G. E. Rowlands, O. J. Lee, D. C. Ralph, and R. A. Buhrman, Appl. Phys. Lett. **105,** 102404 (2014).

137  N. Mikuszeit, O. Boulle, I. M. Miron, K. Garello, P. Gambardella, G. Gaudin, and L. D. Buda-Prejbeanu, Phys. Rev. B **92,** 144424 (2015).

138  C. Zhang, S. Fukami, H. Sato, F. Matsukura, and H. Ohno, Appl. Phys. Lett. **107,** 012401 (2015).

139  N. Perez, E. Martinez, L. Torres, S.-H. Woo, S. Emori, and G. S. D. Beach, Appl. Phys. Lett. **104,** 092403 (2014).

140  E. Martinez, L. Torres, N. Perez, M. A. Hernandez, V. Raposo, and S. Moretti, Sci. Rep. **5,** 10156 (2015).

141  G. Q. Yu, P. Upadhyaya, K. L. Wong, W. J. Jiang, J. G. Alzate, J. S. Tang, P. K. Amiri, and K. L. Wang, Phys. Rev. B **89,** 104421 (2014).

142  G. Finocchio, M. Carpentieri, E. Martinez, and B. Azzerboni, Appl. Phys. Lett. **102,** 212410 (2013).

143  S. Fukami, T. Anekawa, C. Zhang, and H. Ohno, Nat. Nanotechnol. **11,** 621 (2016).

144  M. Baumgartner, K. Garello, J. Mendil, C. O. Avci, E. Grimaldi, C. Murer, J. Feng, M. Gabureac, C. Stamm, Y. Acremann, S. Finizio, S. Wintz, J. Raabe, and P. Gambardella, Nat. Nanotechnol. **12,** 980 (2017).

145  J. Yoon, S.-W. Lee, J. H. Kwon, J. M. Lee, J. Son, X. Qiu, K.-J. Lee, and H. Yang, Sci. Adv. **3,** e1603099 (2017).

146  J. M. Lee, J. H. Kwon, R. Ramaswamy, J. Yoon, J. Son, X. Qiu, R. Mishra, S. Srivastava, K. Cai, and H. Yang, Commun. Phys. **1,** 2 (2018).

147  M. M. Decker, M. S. Wörnle, A. Meisinger, M. Vogel, H. S. Körner, G. Y. Shi, C. Song, M. Kronseder, and C. H. Back, Phys. Rev. Lett. **118,** 257201 (2017).

148  H. Liu, D. Bedau, J. Z. Sun, S. Mangin, E. E. Fullerton, J. A. Katine, and A. D. Kent, J. Magn. Magn. Mater. **358-359,** 233 (2014).

149  D. Bedau, H. Liu, J. Z. Sun, J. A. Katine, E. E. Fullerton, S. Mangin, and A. D. Kent, Appl. Phys. Lett. **97,** 262502 (2010).

150  I. Dzyaloshinskii, JETP **5,** 1259 (1957).

151  T. Moriya, Phys. Rev. **120,** 91 (1960).

152  C. J. Durrant, R. J. Hicken, Q. Hao, and G. Xiao, Phys. Rev. B **93,** 014414 (2016).

153  R. L. Conte, A. Hrabec, A. P. Mihai, T. Schulz, S.-J. Noh, C. H. Marrows, T. A. Moore, and M. Kläui, Appl. Phys. Lett. **105,** 122404 (2014).

154  C. Zhang, S. Fukami, S. DuttaGupta, H. Sato, and H. Ohno, Jpn. J. Appl. Phys. **57,** 04FN02 (2018).

155 G. Yu, P. Upadhyaya, Y. Fan, J. G. Alzate, W. Jiang, K. L. Wong, S. Takei, S. A. Bender, L.-T. Chang, Y. Jiang, M. Lang, J. Tang, Y. Wang, Y. Tserkovnyak, P. K. Amiri, and K. L. Wang, Nat. Nanotechnol. **9,** 548 (2014).

156 G. Yu, M. Akyol, P. Upadhyaya, X. Li, C. He, Y. Fan, M. Montazeri, J. G. Alzate, M. Lang, K. L. Wong, P. Khalili Amiri, and K. L. Wang, Sci. Rep. **6,** 23956 (2016).

157 G. Yu, L.-T. Chang, M. Akyol, P. Upadhyaya, C. He, X. Li, K. L. Wong, P. K. Amiri, and K. L. Wang, Appl. Phys. Lett. **105,** 102411 (2014).

158 M. Akyol, G. Q. Yu, J. G. Alzate, P. Upadhyaya, X. Li, K. L. Wong, A. Ekicibil, P. K. Amiri, and K. L. Wang, Appl. Phys. Lett. **106,** 162409 (2015).

159 C.-F. Pai, M. Mann, A. J. Tan, and G. S. D. Beach, Phys. Rev. B **93**, 144409 (2016).

160 L. You, O. Lee, D. Bhowmik, D. Labanowski, J. Hong, J. Bokor, and S. Salahuddin, Proc. Natl. Acad. Sci. U.S.A. **112,** 10310 (2015).

161 J. Torrejon, F. Garcia-Sanchez, T. Taniguchi, J. Sinha, S. Mitani, J.-V. Kim, and M. Hayashi, Phys. Rev. B **91,** 214434 (2015).

162 W. A. Borders, S. Fukami, and H. Ohno, IEEE Trans. Magn. **53,** 1 (2017).

163 J.-Y. Chen, M. DC, D. Zhang, Z. Zhao, M. Li, and J.-P. Wang, Appl. Phys. Lett. **111,** 012402 (2017).

164 Y.-W. Oh, S.-h. Chris Baek, Y. M. Kim, H. Y. Lee, K.-D. Lee, C.-G. Yang, E.-S. Park, K.-S. Lee, K.-W. Kim, G. Go, J.-R. Jeong, B.-C. Min, H.-W. Lee, K.-J. Lee, and B.-G. Park, Nat. Nanotechnol. **11,** 878 (2016).

165 A. van den Brink, G. Vermijs, A. Solignac, J. Koo, J. T. Kohlhepp, H. J. M. Swagten, and B. Koopmans, Nat. Commun. **7,** 10854 (2016).

166 W. J. Kong, Y. R. Ji, X. Zhang, H. Wu, Q. T. Zhang, Z. H. Yuan, C. H. Wan, X. F. Han, T. Yu, K. Fukuda, H. Naganuma, and M.-J. Tung, Appl. Phys. Lett. **109,** 132402 (2016).

167 Y.-C. Lau, D. Betto, K. Rode, J. M. D. Coey, and P. Stamenov, Nat. Nanotechnol. **11,** 758 (2016).

168 W. Y. Kwak, J. H. Kwon, P. Grünberg, S. H. Han, and B. K. Cho, Sci. Rep. **8,** 3826 (2018).

169 S. A. Razavi, D. Wu, G. Yu, Y.-C. Lau, K. L. Wong, W. Zhu, C. He, Z. Zhang, J. M. D. Coey, P. Stamenov, P. Khalili Amiri, and K. L. Wang, Phys. Rev. Appl. **7,** 024023 (2017).

170 Z. Zhao, A. K. Smith, M. Jamali, and J.-P. Wang, "External-Field-Free Spin Hall Switching of Perpendicular Magnetic Nanopillar with a Dipole-Coupled Composite Structure", preprint arXiv:1603.09624 (2017 ).

171 K. Cai, M. Yang, H. Ju, S. Wang, Y. Ji, B. Li, K. W. Edmonds, Y. Sheng, B. Zhang, N. Zhang, S. Liu, H. Zheng, and K. Wang, Nat. Mater. **16,** 712 (2017).

172 V. P. Amin and M. D. Stiles, Phys. Rev. B **94,** 104419 (2016).

173 V. P. Amin and M. D. Stiles, Phys. Rev. B **94,** 104420 (2016).

174 A. M. Humphries, T. Wang, E. R. J. Edwards, S. R. Allen, J. M. Shaw, H. T. Nembach, J. Q. Xiao, T. J. Silva, and X. Fan, Nat. Commun. **8,** 911 (2017).

175 S.-h. C. Baek, V. P. Amin, Y.-W. Oh, G. Go, S.-J. Lee, G.-H. Lee, K.-J. Kim, M. D. Stiles, B.-G. Park, and K.-J. Lee, Nat. Mater. **17,** 509 (2018).

176 J. M. Lee, K. Cai, G. Yang, Y. Liu, R. Ramaswamy, P. He, and H. Yang, Nano Lett.**,** preprint doi:10.1021/acs.nanolett.8b00773 (2018).